\documentclass[nofootinbib,twocolumn,superscriptaddress,
amsmath,amssymb,aps,pra]{revtex4-1}

\usepackage{graphicx}
\usepackage{physics}
\usepackage{comment}
\usepackage{color}
\usepackage{dcolumn}
\usepackage[normalem]{ulem}
\usepackage{bm}
\usepackage[colorlinks,urlcolor=blue,citecolor=blue,linkcolor=blue]{hyperref}

\newcommand{\vect}[1]{{\mathbf #1}}
\newcommand{\vegr}[1]{{\boldsymbol #1}}   
\renewcommand{\k}{{\bf k}}

\newcommand{\0}{{\bf 0}}

\newcommand{\beq}{\begin{equation}}
\newcommand{\eeq}{\end{equation}}

\newcommand{\area}{\mathcal{A}}
\newcommand{\sch}{Schr{\"o}dinger }
\renewcommand{\r}{{\bf r}}

\newcommand{\Frac}[2]{\displaystyle\frac{#1}{#2}}

\begin{document}

\title{Rydberg excitons and polaritons in monolayer transition metal dichalcogenides in a magnetic field}

\author{David de la Fuente Pico}
\affiliation{Departamento de F\'isica Te\'orica de la Materia
  Condensada, Universidad
  Aut\'onoma de Madrid, Madrid 28049, Spain}
\affiliation{Condensed Matter Physics Center (IFIMAC), Universidad Autónoma de Madrid, 28049 Madrid, Spain}

\author{Jesper Levinsen}
\affiliation{School of Physics and Astronomy, Monash University, Victoria 3800, Australia}
\affiliation{ARC Centre of Excellence in Future Low-Energy Electronics Technologies, Monash University, Victoria 3800, Australia}

\author{Emma Laird}

\affiliation{ARC Centre of Excellence in Future Low-Energy Electronics and Technologies, University of
Queensland, Queensland 4072, Australia}

\author{Meera M. Parish}
\affiliation{School of Physics and Astronomy, Monash University, Victoria 3800, Australia}
\affiliation{ARC Centre of Excellence in Future Low-Energy Electronics Technologies, Monash University, Victoria 3800, Australia}

\author{Francesca Maria Marchetti}
\affiliation{Departamento de F\'isica Te\'orica de la Materia
  Condensada, Universidad
  Aut\'onoma de Madrid, Madrid 28049, Spain}
\affiliation{Condensed Matter Physics Center (IFIMAC), Universidad Autónoma de Madrid, 28049 Madrid, Spain}
  
\date{January 23, 2025}

\begin{abstract}
We develop a microscopic theory for excitons and cavity exciton polaritons in transition metal dichalcogenide (TMD) monolayers under a perpendicular static magnetic field. 
We obtain numerically exact solutions for the ground and excited states, accounting for the interplay between arbitrarily large magnetic fields and light-matter coupling strengths. This includes the very strong coupling regime, where light-induced modifications of the exciton wavefunction become essential and the approximate coupled oscillator description breaks down. 
Our results show excellent agreement with recent experimental measurements of the diamagnetic shift of the ground and excited exciton states in WS$_2$, MoS$_2$, MoSe$_2$, and MoTe$_2$ monolayers. For polaritons, we consider experimentally relevant system parameters and demonstrate that the diamagnetic shifts of both the ground and excited states at high magnetic fields exhibit clear signatures of the very strong coupling regime, highlighting the necessity of our microscopic and numerically exact treatment over perturbative approaches.
Furthermore, our microscopic approach allows us to evaluate the exciton-exciton and polariton-polariton interaction strengths. Comparing results specific to TMD monolayers with those applicable to quantum wells, we find that variational approaches overestimate the TMD excitons' interaction strength. We also observe that magnetic fields weaken the interaction strength for both excitons and polaritons, with a less pronounced effect in TMDs than in quantum wells, and that light-induced modifications to the matter component in TMD polaritons can enhance interaction strengths beyond those of purely excitonic interactions.
\end{abstract}

\maketitle
\section{Introduction}
\label{sec:intro}
Magneto-optical spectroscopy plays an essential role in investigating two-dimensional (2D) semiconductors~\cite{Miura_book2007} and is highly effective for manipulating and exploring the properties of excitons.
The application of this technique to traditional 2D semiconductors, such as III-V~\cite{Tarucha_SSC1984,Bugajski_SSC1986} and II-VI~\cite{Warnock_PRB1985,Ivchenko_PRB1992} quantum wells (QWs), has a long history, where it has been instrumental in manipulating and analyzing exciton properties. In this context, a magnetic field in a perpendicular geometry has been used to distinguish and identify both the ground and excited exciton states. In addition, the resulting diamagnetic shifts provide valuable information about the exciton reduced mass, size, and spin, enabling a direct comparison of experimental results with Wannier-based exciton models~\cite{MacDonald-Ritchie_PRB1986,Edelstein_PRB1989,Stafford_PRB1990}.

The magnetic field acts as an in-plane confining potential, further binding the excitons and enhancing their coupling to light. This effect has been employed to strongly couple the ground and excited exciton states to light to form cavity exciton polaritons (polaritons for short) when the QW is embedded into a microcavity~\cite{Tignon_PRL1995,Pietka_PRB2015,Pietka-Potemski_PRB2017,Brodbeck_PRL2017}. Indeed, for III-V heterostructures, the Rabi couplings between the excited 
exciton states and a microcavity photon are hardly detectable at zero magnetic field, but they are enhanced when a magnetic field is applied~\cite{Whittaker_NuovoCimentoD1995,Fisher-Whittaker_PRB1996,Berger_PRB1996} and can therefore be measured.
Importantly, the diamagnetic shift of polariton modes has been proposed~\cite{Yang-Yamamoto_NJP2015} and utilized~\cite{Brodbeck_PRL2017,Laird_PRB2022} as a method to confirm the realization of the very strong light-matter coupling regime. In this regime, the Rabi coupling approaches the exciton binding energy, resulting in the hybridization of different excitonic states within a single polariton state. 

More recently, significant advances have been made in the study of atomically thin 2D semiconductors. In particular, transition metal dichalcogenide (TMD) monolayers are emerging as unique 2D optically active materials with remarkable properties arising from their low dimensionality and unique band structure, opening new directions of study in optoelectronics, as well as complementing those for traditional quantum well structures~\cite{Wang_RMP2018}.
TMD monolayers have a direct bandgap together with valley- and spin-dependent selection rules, leading to potential applications in valleytronics, optoelectronics and the fabrication of nanophotonic devices~\cite{Schaibley_NatRevMat2016,Mak_NatPhot2016,Liu_NanoRes2019}. 
Furthermore, they exhibit pronounced exciton resonances even at room temperature. 
Indeed, due to heavy carrier masses and reduced dielectric screening, excitons in TMD monolayers are characterized by very  large binding energies, on the order of hundreds of meV~\cite{Rasumbraniam_PRB2012,Berkelbach_PRB2013,Chernikov_PRL2014}. 
Their strong coupling to light led to the first observation of room-temperature polaritons with
TMD monolayers embedded in a microcavity~\cite{Liu-Menon_NatPhot2015,Dufferwiel_NatCom2015}, and to the first observation of  polariton states with excited exciton states~\cite{Gu-Menon_NatComm2021}. 
Similar to QWs, stacking multiple TMD monolayers  inside an optical microcavity leads to an increase in the Rabi coupling towards the very strong coupling regime~\cite{Zhao-Sanvitto_NatComm2023}.

Magneto-optical measurements have already been employed for the quantitative analysis of TMD monolayer parameters and for gaining detailed insights into spin-valley physics — for a review, see Ref.~\cite{Arora_JAppPhys2021}.
As with III-V and II-VI quantum wells, exciton properties in TMD monolayers --- such as their reduced mass, size, binding energy, and dependence on the dielectric environment --- have been studied using magnetic fields~\cite{Mitioglu_NanoLett2015,Plechinger_NanoLett2016,Stier-Crooker_NanoLett2016,Stier-Crooker_NatComm2016,Stier-Crooker_PRL2018,Zipfel-Chernikov-MF_PRB2018,Have_PRB2019,Liu_PRB2019,Delhomme_APL2019,Goryca-Crooker_NatCom2019,Molas-Potemski_PRL2019}.
The main differences for TMD monolayers in comparison to III-V and II-VI quantum wells are the much larger exciton binding energies and smaller radii, which require much stronger magnetic fields --- on the order of tens of Tesla --- to observe a significant diamagnetic shift in the exciton ground state. However, the higher exciton oscillator strengths facilitate the detection of excited states. 
In TMD monolayers such as MoS$_2$, MoSe$_2$, MoTe$_2$, and
WS$_2$, high magnetic fields of up to $91$~T~\cite{Goryca-Crooker_NatCom2019}  have made it possible to observe the first five Rydberg exciton states.
The distinct shifts of the excited states provide a direct means to quantitatively compare experimental data with theoretical models --- see Refs.~\cite{Stier-Crooker_PRL2018,Goryca-Crooker_NatCom2019}.

In this paper, we develop a microscopic theory for excitons in a TMD monolayer under a perpendicular static magnetic field, as well as for polaritons when one or several monolayers are embedded in a microcavity.
Numerically exact solutions are obtained for the ground and excited states for arbitrary large magnetic fields and light-matter coupling strengths, extending to the very strong coupling regime, where it is crucial to take light-induced modifications of the exciton wavefunction 
into account.
Generalizing previous results valid for quantum wells~\cite{Laird_PRB2022}, the numerically exact solution takes advantage of a mapping between the 2D hydrogen atom and the 2D harmonic oscillator.
We find excellent agreement between our results and experimental measurements of the diamagnetic shift of the ground and excited exciton states for 
MoS$_2$, MoSe$_2$, MoTe$_2$, and
WS$_2$ monolayers at high fields --- 
Note that in Refs.~\cite{Stier-Crooker_PRL2018,Goryca-Crooker_NatCom2019}, numerically exact solutions for the diamagnetic shifts at arbitrary magnetic field were already achieved employing a numerical procedure different from the one discussed here. 
For polaritons we consider system parameters accessible to current experiments~\cite{Zhao-Sanvitto_NatComm2023}. We demonstrate that the diamagnetic shift of the ground and excited states at large values of the magnetic field carries signatures of the very strong coupling regime between light and matter, which  strongly differentiate our exact results from perturbative ones.

We furthermore employ a Born approximation to estimate the ground-state exciton-exciton and polariton-polariton interaction strengths and their dependence on the magnetic field. 
We evaluate these  
for TMD monolayers and for typical III-V quantum wells, thus allowing us to directly compare these two platforms. 
We find that the magnetic field decreases the interaction strength both between excitons and between polaritons, with a more pronounced effect for QWs compared to TMD monolayers, since the latter have substantially larger exciton binding energies. 
For TMD excitons, we show that the use of hydrogenic variational wavefunctions~\cite{Shahnazaryan_PRB_2017} overestimates the interaction strength, leading to qualitatively incorrect results and highlighting the need for exact solutions.
Finally, we demonstrate for TMD polaritons that incorporating light-induced modifications to the matter component can lead to interaction strengths that exceed those of purely excitonic interactions.

The paper is organized as follows: In Sec.~\ref{sec:model}, we present the microscopic model describing TMD excitons and polaritons in a static magnetic field.
The formalism required to obtain numerically exact solutions for the full Rydberg series of magnetoexcitons and magnetopolaritons is presented in Sec.~\ref{sec:excitons} and Sec.~\ref{sec:polaritons}, respectively. The results, along with comparisons to experimental data and predictions for future studies, are discussed in these respective sections. In Sec.~\ref{sec:X-inter_strength} and Sec.~\ref{sec:inter_strength}, we evaluate the exciton-exciton and polariton-polariton interaction strengths, respectively, and we discuss the effects of both a strong magnetic field and very strong coupling to light. Conclusions are drawn in Sec.~\ref{sec:conclusions}.

\begin{figure}
    \includegraphics[width=0.8\columnwidth]{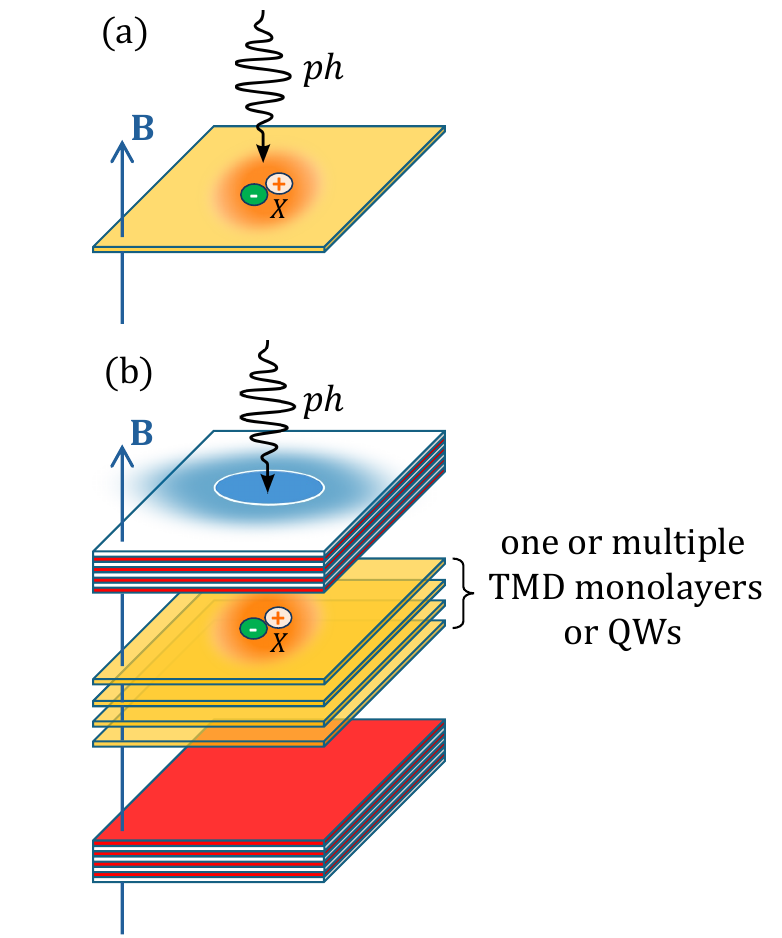}
    \caption{Schematic representation of the two set-ups considered in this work. (a) A TMD monolayer or single quantum well in a perpendicular homogeneous magnetic field $\textbf{B}$ is optically excited by a resonant circularly polarized photon ($ph$) generating an electron-hole pair which form an exciton ($X$). (b) One or several encapsulated TMD monolayers or QWs in a magnetic field are embedded into an optical microcavity excited at normal incidence.}
    \label{fig:schematic}
\end{figure}
\section{Model}
\label{sec:model}
In this section, we introduce the microscopic model employed to describe conduction electron and valence hole charges in a TMD monolayer that can form excitons.  Similar to previous work~\cite{Stier-Crooker_PRL2018,Goryca-Crooker_NatCom2019}, our 
model allows us to capture non-perturbative effects arising from a strong perpendicular magnetic field in TMD monolayers. 
In addition, we include the possibility of describing cavity photons and their coupling to matter excitations when the monolayer is 
embedded in an optical microcavity (see Fig.~\ref{fig:schematic}). 
Further, in this work we assume that electron-hole pairs are 
excited across the energy gap of a single valley by circularly polarized light, so that excitons are spin-valley polarized and we can formally neglect these degrees of freedom. Since our focus is on bright excitons that are coupled to light, we do not consider the exciton fine structure involving optically dark states.

The system Hamiltonian includes three terms: $\hat{H}_{m}$ depicts the matter-only excitations, i.e., conduction electrons and valence holes,  $\hat{H}_{C}$ the cavity photon, and $\hat{H}_{Cm}$ the interaction between light and matter: 
\begin{equation}
\label{eq:Hamil_all_terms}
    \hat{H}=\hat{H}_{m}+\hat{H}_{C}+\hat{H}_{Cm}\; .
\end{equation}
In the presence of a magnetic field, the electrons and holes in a TMD monolayer are described by the Hamiltonian (throughout this work we set $\hbar=1$)
\begin{widetext}
\begin{equation}
\label{eq:Hamil_matter}
    \hat{H}_{m}= \sum_{\sigma=e,h}\int d\r \hat{\Psi}_\sigma^{\dag}(\r) \frac{[-i\nabla \pm \frac{e}{c} \vect{A}(\r)]^2}{2m_\sigma} \hat{\Psi}_\sigma (\r) + \Frac{1}{2} \sum_{\sigma, \sigma'} \int d\r_1 d\r_2 \hat{\Psi}^{\dag}_\sigma(\r_1) \hat{\Psi}^{\dag}_{\sigma'}(\r_2) W_{\sigma \sigma'}(\r_1-\r_2)\hat{\Psi}_{\sigma'}(\r_2)\hat{\Psi}_\sigma(\r_1)\; ,
\end{equation}
\end{widetext}
where $\hat{\Psi}_\sigma^\dag (\r)$ ($\hat{\Psi}_\sigma^{} (\r)$) are the creation (annihilation) field operators for electrons and holes ($\sigma=e, h$) at the position $\r$, $m_{e,h}$ their effective masses, and the sign $+$ ($-$) is for electrons (holes).\footnote{Strong experimental support for the applicability of the Wannier-Mott model and effective mass theory to describe the key properties of excitons in TMD monolayers first came from studies of exciton radii in 
WS$_2$ monolayers using magneto-optical reflectance spectroscopy~\cite{Zipfel-Chernikov-MF_PRB2018}.} 
Note that, unless mentioned explicitly, throughout this work we measure
the energies with respect to the bandgap energy $E_g$. 
The operators satisfy canonical anticommutation relations:
\begin{subequations}
\begin{align}
    \{\hat{\Psi}_{\sigma}^{}(\r),\hat{\Psi}_{\sigma'}^{\dag}(\r')\}&=\delta(\r-\r')\delta_{\sigma\sigma'}\\
    \{\hat{\Psi}_{\sigma}^{}(\r),\hat{\Psi}_{\sigma'}^{}(\r')\} &= \{\hat{\Psi}_{\sigma}^{\dag}(\r),\hat{\Psi}_{\sigma'}^{\dag}(\r')\} = 0\; .
\end{align}    
\end{subequations}
In view of the fact that, later on, we are interested in evaluating polariton-polariton interaction properties, we find it convenient to work with a second quantization formalism in the real space representation. The model can equivalently be formulated within momentum space, see Ref.~\cite{Laird_PRB2022} for details.

We assume that the magnetic field  $\mathbf{B}=\nabla\times\mathbf{A}(\r)=(0,0,B)$ is uniform and oriented in the  
$z$ direction, perpendicularly to the 
$x$-$y$ plane of the TMD monolayer. We 
work in the symmetric gauge, where the vector potential is given by:
\begin{equation}
\label{eq:symmetric_gauge}
    \mathbf{A} (\r)=\frac{1}{2}\mathbf{B}\times\r =\Frac{B}{2} 
        (-y, x, 0)
    \; .
\end{equation}
Note that this gauge choice is 
consistent with the Coulomb gauge $\nabla \cdot \mathbf{A(\r)}=0$, employed when describing the coupling to light. 

The last term in Eq.~\eqref{eq:Hamil_matter} depicts the attractive interaction between an electron and a hole $W_{eh}(\r)$, leading to the formation of excitons, as well as the repulsion either between two electrons $W_{ee}(\r)$ or two holes $W_{hh}(\r)$ when multiple electron-hole pairs are photo-excited: 
\begin{subequations}
\label{eq:interactions_eh_ee_hh}
\begin{align}
    W_{eh}(\r)=&V (r)\\
    W_{ee}(\r)=&W_{hh}(\r)=-V(r)\; .
\end{align}
\end{subequations}
For TMD monolayers, either encapsulated within two dielectrics or held in vacuum, the electron-hole interaction has been shown to be appropriately accounted for by the Rytova-Keldysh potential~\cite{Rytova_Moscow1967,Keldysh_SovietJournal1979} (throughout this work we use Gaussian units $4\pi\epsilon_0=1$, where $\epsilon_0$ is the vacuum permittivity):
\begin{equation}
\label{eq:Rytova-Keldysh-potential}
     V (r)=-\frac{\pi e^2}{2 r_0 \varepsilon}\left[H_0\left(\frac{r}{r_0}\right)-Y_0\left(\frac{r}{r_0}\right)\right]\; .
\end{equation}
Here, $H_0(x)$ is the zeroth order Struve function and $Y_0(x)$ is the zeroth-order Bessel function of the second kind. TMD monolayers are typically encapsulated, often with hexagonal boron nitride (hBN), which significantly enhances their optical quality~\cite{Wierzbowski_SciReports2017,Cadiz_PRX2017}. 
In Eq.~\eqref{eq:Rytova-Keldysh-potential}, the influence of the surrounding materials, characterized by dielectric constants $\varepsilon_{1,2}$, is modelled by introducing an effective dielectric constant $\varepsilon=\frac{1}{2}(\varepsilon_{1}+\varepsilon_{2})$, while $r_0$ is the screening length. For a TMD monolayer having thickness $d$ and dielectric constant $\varepsilon_{ML}$, the screening length is defined as $r_0=d \varepsilon_{ML}/(2\varepsilon) $~\cite{Berkelbach_PRB2013}. In the strictly 2D limit ($d\rightarrow0)$ and in vacuum ($\varepsilon=1$), the screening length is defined in terms of the 2D polarizability of the monolayer $\chi_{2D}$, as   $r_0=2\pi\chi_{2D}$~\cite{Cudazzo_PRB2011}. 
The screening length $r_0$ marks the crossover between short- and long-range behavior:
At short distances, the Rytova-Keldysh potential diverges as $\sim \ln r$, accounting for the confinement of field lines in the 2D plane. Instead, at large distances, it recovers the Coulomb potential dependence $\sim r^{-1}$. The non-local screening introduced by this potential has been shown to accurately describe exciton properties in TMD monolayers~\cite{Chernikov_PRL2014,Cudazzo_PRB2011,Berkelbach_PRB2013}. Furthermore, the Rytova-Keldysh potential is also commonly used to evaluate electron-electron and hole-hole interactions. However, its validity for more complex structures, such as trions, is a subject of debate~\cite{VanTuan_PRB2018}.

The photonic part of the Hamiltonian is:
\begin{equation}  
    \hat{H}_{C}= \omega \hat{a}^\dag \hat{a}^{}\; ,
\end{equation}%
where $\omega$ is the bare cavity energy (i.e., the cavity photon energy in the absence of an embedded active medium~\cite{Levinsen_PRR2019}) and $\hat{a}^{\dag}$ $(\hat{a})$ is the creation (annihilation) operator for the cavity photon. For simplicity, we only consider a single cavity mode. This either corresponds to the case of a 0D cavity, such as fiber cavities~\cite{Delteil_NatMat2019},  
or to the case of a photon mode with zero in-plane momentum of a 2D microcavity excited at normal incidence; see Fig.~\ref{fig:schematic}(b). In this case, because of the photon mass being about 5 orders of magnitude smaller than the mass of matter excitations, we can account for the finite photon momentum by simply shifting the cavity photon frequency.

Finally, the Hamiltonian term describing the light-matter coupling is:
\begin{equation}
\label{eq:ham-light-mat-inter}
\hat{H}_{Cm}=\frac{g}{\sqrt{\area}} \left[\hat{a}^{} \int d\r  \hat{\Psi}_{e}^{\dag}(\r)\hat{\Psi}_{h}^{\dag}(\r) +\text{h.c}.\right],
\end{equation}
where $g$ is the light-matter coupling strength and $\area$ the system area.
This term describes 
the creation (annihilation) of an electron-hole pair with zero electron-hole separation by absorption (emission) of a photon. Because of the effects of the active medium, approximating this interaction as effectively having zero range implies that the bare cavity energy $\omega$ needs to be renormalized by introducing the dressed photon energy~\cite{Levinsen_PRR2019}, as discussed in detail in Sec.~\ref{sec:polaritons}. 
Equation \eqref{eq:ham-light-mat-inter} allows us to 
describe very strong light-matter effects leading to the hybridization of different exciton states in the formation of  polaritons. However, we assume that the TMD bandgap energy is much larger than the light-matter coupling strength, such that we can work in the rotating-wave approximation. This means that we do not describe the regime of ultrastrong coupling, where hybridization with different numbers of excitations occurs.

\section{Magnetoexcitons}
\label{sec:excitons}
In this section, we describe the formalism necessary to study the properties of excitons in TMD monolayers in a perpendicular magnetic field. 
A numerically exact microscopic theory of quantum well excitons in a perpendicular magnetic field was  already developed in Ref.~\cite{Laird_PRB2022} by making use of an exact mapping between the 2D harmonic oscillator and the 2D hydrogen atom. It was shown that this allowed a very efficient numerical solution of the problem, for any strength of the magnetic field. Here, we generalize those results valid for Coulomb interactions to the case of a Rytova-Keldysh potential, 
thus allowing us to describe current experiments on TMD monolayers. In particular, we demonstrate that our theory agrees very well with the exciton diamagnetic shift measured in recent magneto-optical experiments in Ref.~\cite{Goryca-Crooker_NatCom2019}. Note that the Schrödinger equation describing $s$-wave TMD magnetoexcitons was solved in real space in Refs.~\cite{Stier-Crooker_PRL2018, Goryca-Crooker_NatCom2019} for arbitrary magnetic field strengths, showing excellent agreement with experiments. However, as discussed next, we employ a different numerical approach that avoids the manipulation of derivatives.  
We first review the formal steps followed to exactly solve the problem numerically, including the possibility of describing states with any orbital angular momentum, and the necessary adaptation to the case of TMD monolayers.

In the presence of a homogeneous perpendicular magnetic field, the translational symmetry is preserved while the total linear momentum of an electron-hole pair is no longer conserved. Instead, the conserved quantity is the total magnetic momentum $\widehat{\mathbf{K}}$ (the generator of translations), defined as
\begin{multline}
    \widehat{\vect{K}}=-i\nabla_{\r_e}-\frac{e}{c}\vect{A}(\r_e)-i\nabla_{\r_h}+\frac{e}{c}\vect{A}(\r_h)\\
    =-i\nabla_{\vect{R}}-\frac{e}{2c}\vect{B}\times \r\; ,
\end{multline}
where 
\begin{subequations}
\begin{align}
    \vect{R} &=\frac{m_e\r_e+m_h\r_h}{m_e+m_h}\\
    \r &=\r_e-\r_h\; ,
\end{align}
\end{subequations}
are the center of mass and relative positions, respectively.
Because of this, the most general single exciton state in the center of mass frame can be written as
\begin{equation}
\label{eq:exciton-state}
    |X_{\vect{K}}^{} \rangle = \hat{X}_{\vect{K}}^\dag \ket{0}\; ,
\end{equation}
where the creation operator of an exciton with total magnetic momentum $\vect{K}$  is
\begin{multline}
\label{eq:Xoperator}
    \hat{X}_{\vect{K}}^\dag = \frac{1}{\sqrt{\area}}\int d\r_e d\r_h e^{i\left(\mathbf{K}+\frac{e}{2c}\mathbf{B}\times\r\right)\cdot \vect{R}}\\  
    \times \varphi_{\mathbf{K}}(\r)\hat{\Psi}_e^{\dag}(\mathbf{r}_e)\hat{\Psi}_h^{\dag}(\mathbf{r}_h)\; ,
\end{multline}
where $\varphi_{\mathbf{K}}(\r)$ is the wavefunction  describing the relative motion of the electron and hole. Equation~\eqref{eq:Xoperator} is referred to as the Lamb transformation and it was introduced first by Lamb in Ref.~\cite{Lamb_PR1952} for the hydrogen atom and later adapted to the exciton problem by Gor’kov and Dzyaloshinskii in Ref.~\cite{Gorkov_SovietJourn1968}. 

In order to derive the \sch equation satisfied by the exciton wavefunction, we first evaluate the expectation value of the matter Hamiltonian~\eqref{eq:Hamil_matter} for the state~\eqref{eq:exciton-state}:
\begin{equation}
    \langle X_{\vect{K}}^{} | \hat{H}_m|X_{\vect{K}}^{} \rangle=\int d\r \varphi^{*}_{\mathbf{K}}(\r)\widehat{H}_m'\varphi_{\mathbf{K}}(\r)\; ,
\end{equation}
with
\begin{multline}\label{eq:trans-mat-hamiltonian}
    \widehat{H}_m'= \left[-\frac{\nabla_{\r}^2}{2\mu} -i \frac{e\eta}{2\mu c} \mathbf{B}\cdot (\r\times \nabla_{\r})+\frac{e^2}{8\mu c^2}(\mathbf{B}\times\r)^2 \right.\\
    \left. +\frac{e}{Mc}(\vect{K}\times\mathbf{B})\cdot \r+\frac{\mathbf{K}^2}{2M}+V (r)
    \right]\; ,
\end{multline}
where $\mu=\frac{m_e m_h}{m_e+m_h}$ and $M=m_e+m_h$ are the exciton reduced and total masses, respectively, while $\eta=(m_h-m_e)/(m_e+m_h)$. The first and second terms of this expression can be rewritten in terms of the $z$-component of the relative orbital angular momentum operator $\widehat{L}_z = -i x \partial_y + i y \partial_x = -i \partial_\theta $ (with eigenvalue $l_z$):
\begin{subequations}
\label{eq:Lz}
\begin{align}
\label{eq:Lz_1}  
    -i \frac{e\eta}{2\mu c} \mathbf{B}\cdot (\r\times \nabla_{\r}) &=\frac{\omega_{c,e}-\omega_{c,h}}{2}\widehat{L}_z \\
    -\nabla_{\r}^2 &=-\partial_{r}^2-\frac{1}{r}\partial_r+\frac{\widehat{L}^2_z}{r^2}\;,
\label{eq:Lz_2}    
\end{align}
\end{subequations}
where $\omega_{c,j}=\frac{eB}{m_j c}$ ($j=e,h$) are the electron and hole cyclotron frequencies. 
Equation~\eqref{eq:Lz_1} corresponds to a simple Zeeman shift which only affects the exciton's energy and not the exciton wavefunction, since orbital angular momentum $l_z$ is a good quantum number.  
In addition, we focus on $s$-wave exciton states where $l_z = 0$ in~\eqref{eq:Lz}, since one-photon transitions in TMD monolayers only allow the excitations of isotropic exciton states, i.e., those  with $s$-wave symmetry and zero relative orbital angular momentum. Finite orbital angular momentum excitonic states are also referred to as `dark' and cannot be accessed with linear optics. However, two-photon transitions can allow one to access dark excitons with odd parity~\cite{Ye_Nature2014}. Dark excitons can alternatively be detected by the application of a weak static in-plane electric field, which can induce orbital hybridization between Rydberg excitonic states with different angular momenta~\cite{PhysRevLett.131.036901}. We investigate the extent to which our theoretical approach can be utilized to study excitons with finite orbital angular momentum in Appendix~\ref{app:finiteLz}.

A normally incident photon generates an electron-hole pair with zero center of mass momentum and zero separation, i.e, with magnetic momentum $\vect{K}=0$. Since the magnetic momentum is a good quantum number, we can set it to zero from now on. In this case, the \sch equation for the relative exciton wavefunction $\varphi_{\0} (\r) \equiv \varphi (r)$ reads as:
\begin{multline}
\label{eq:sch-real-space-exciton}
    E\varphi(r)=\\\left[-\frac{1}{2\mu} \left(\frac{d^2}{dr^2}+\frac{1}{r}\frac{d}{dr}\right) + \frac{\mu \omega_c^2}{2}r^2+V  (r)\right]\varphi(r)\;,
\end{multline}
where $\omega_c=\frac{eB}{2\mu c}$ is the exciton cyclotron frequency. 

The \sch equation~\eqref{eq:sch-real-space-exciton} can be solved perturbatively in the two regimes of weak and strong magnetic field. 
In the weak magnetic field limit, i.e., when $\omega_c\ll |E_{ns}^{B=0}|$, where $|E_{ns}^{B=0}|$ is the $ns$ exciton binding energy for zero magnetic field, the perturbative expression for the exciton energy reads as:
\begin{equation}
\label{eq:perturbative_smallB}
    E_{ns} \simeq E_{ns}^{B=0} + \frac{\mu\omega_c^2}{2}\langle r^2\rangle_{ns}\; ,
\end{equation}
where $\langle r^2\rangle_{ns}$ is the squared mean radius of the exciton, evaluated in the zero magnetic field limit. The electron-hole relative wavefunctions can also be evaluated perturbatively in this limit as:
\begin{multline}
\label{eq:wf_perturbative_smallB}
    \varphi_{ns}(r)\simeq \varphi_{ns}^{B=0} (r)\\
    +\frac{\mu \omega_c^2}{2}\sum_{m\neq n} \frac{\langle m | r^2 | n \rangle}{E^{B=0}_{ns}-E_{ms}^{B=0}} \varphi_{ms}^{B=0}(r)\; ,
\end{multline}
where $\langle m | r^2 | n \rangle = \int d\r \varphi_{ms}^{B=0 *} (r) r^2  \varphi_{ns}^{B=0} (r)$. 
In the opposite limit of a strong magnetic field, when $\omega_c \gg |E_{ns}^{B=0}|$, one instead has~\cite{MacDonald-Ritchie_PRB1986}
\begin{equation}
\label{eq:perturbative_largeB}
    E_{ns} \simeq \omega_c\left[(2n-1)+ \mathcal{O}\left(\frac{1}{\sqrt{\omega_c}}\right)\right]\; .
\end{equation}

In the regime of intermediate magnetic field strengths, Eq.~\eqref{eq:sch-real-space-exciton} has to be solved numerically. This can be carried out efficiently by mapping the 2D harmonic oscillator problem into the 2D hydrogen problem, as already discussed in Refs.~\cite{Laird_PRB2022,Duru_ForsDerPhysik1982}. We thus perform the change of variables $\rho=\frac{r^2}{8a_X^2}$ and $\bar{E} = E/R_X$, where $R_X$ and $a_X$
are the exciton Rydberg and Bohr radius, 
\begin{align}
\label{eq:hydrogen_scales}
    R_X &=\frac{2\mu e^4}{\varepsilon^2}=\frac{1}{2\mu a_X^2} & a_X &=\frac{\varepsilon}{2\mu e^2}\; .
\end{align}
These are the energy and length scales characterizing the 2D hydrogenic exciton problem for pure Coulomb interaction~\cite{klingshirn2012semiconductor}, i.e., $R_X = \lim_{r_0 \to 0}|E_{1s}^{B=0}|$. The change of variables results in the following dimensionless equation:
\begin{equation}
\label{eq:rescaled}
    \frac{2 \bar{E}}{\rho}\bar{\varphi}(\vegr{\rho})=\left[-\Frac{d^2}{d\rho^2} - \Frac{1}{\rho} \Frac{d}{d\rho}  + 4\bar{\omega}_c^2+\tilde{V} (\rho) \right]\bar{\varphi}(\vegr{\rho})\; ,
\end{equation}
where $\bar{\varphi}(\vegr{\rho})=a_X \varphi (\r)$, $\bar{\omega}_c=\omega_c/R_X$, and $\tilde{V} (\rho)= \frac{2V (\sqrt{8a_X^2\rho})}{R_X\rho}$. 
We can now carry out the Fourier transform from the rescaled real coordinate $\rho$ to the rescaled reciprocal space $\vegr{\kappa}$, 
by applying 
$\int d\bm{\rho} e^{-i\bm{\kappa}\cdot\bm{\rho}}\{ \cdot \}$ to both sides of Eq.~\eqref{eq:rescaled}, obtaining:
\begin{equation}
\label{eq:sch-eq-kappa}
    \bar{E}\sum_{\bm{\kappa}'}\frac{4\pi\bar{\varphi}_{\kappa'}}{|\bm{\kappa}-\bm{\kappa}'|} = (\kappa^2 + 4\bar{\omega}_c^2)\bar{\varphi}_{\kappa}+\sum_{\bm{\kappa}'} \tilde{V} _{|\bm{\kappa}-\bm{\kappa}'|}\bar{\varphi}_{\kappa'}\; ,
\end{equation}
where $\vegr{\kappa} = (\kappa, \phi)$, we use the notation $\sum_{\bm{\kappa}}\equiv\int\frac{d\bm{\kappa}}{(2\pi)^2} = \int_0^\infty \frac{d\kappa \kappa}{2\pi} \int_0^{2\pi} \frac{d\phi}{2\pi}$, and $\tilde{V}_{\kappa}=\int d\bm{{\rho}}  e^{-i\vegr{\kappa}\cdot \bm{\rho}}\tilde{V} (\rho)$.
The analytic expression for $\tilde{V}_{\kappa}$ is given in Eq.~\eqref{eq:RK-kappa-explicit}.
Note that the previously obtained eigenfunctions must be properly normalized in order to ensure that
\begin{multline}
\label{eq:normalization}
    1=\int d\r|\varphi(r)|^2=\int d\bm{\rho} \frac{4}{\rho}|\bar{\varphi}(\rho)|^2 \\
    =8 \pi \sum_{\bm\kappa,\bm\kappa'}\frac{\bar{\varphi}_{\kappa} \bar{\varphi}_{\kappa'}^{*}}{|\bm{\kappa}-\bm{\kappa}'|}\;.
\end{multline}
This can easily be achieved if we take $\bar{\varphi}_{\kappa}\rightarrow \bar{\varphi}_{\kappa}/\mathcal{N}$ with $\mathcal{N}^2=8\pi\sum_{\bm{\kappa},\bm{\kappa'}}\frac{\bar{\varphi}_{\kappa} \bar{\varphi}_{\kappa'}^{*}}{|\bm{\kappa}-\bm{\kappa}'|}$. 
This change is implicitly assumed throughout the remainder of this section.

The \sch equation in rescaled momentum $\kappa$ space~\eqref{eq:sch-eq-kappa} can be readily solved by direct diagonalization once the rescaled momentum $\kappa$ is discretized on a grid. In particular, we choose to use 
a Gauss-Legendre quadrature ~\cite{Press_Numerical_recipes2007}. 
Both the Rytova-Keldysh term 
$\tilde{V}_{|\vegr{\kappa}-\vegr{\kappa}'|}$, as well as the 2D Coulomb-like term $\frac{4\pi}{|\bm{\kappa}-\bm{\kappa}'|}$ have a pole for $\bm{\kappa}=\bm{\kappa}'$. In order to deal with these singularities, we implement the subtraction scheme proposed in Ref~\cite{Laird_PRB2022}. This subtraction scheme allows us to cancel the singularities at $\bm{\kappa}=\bm{\kappa}'$ in both terms,\footnote{Note that the subtraction scheme should also be applied to evaluate the normalization constant, $\mathcal{N}^2$, below Eq.~\eqref{eq:normalization}.} without fictitiously removing them, thus significantly accelerating the numerical convergence. Details are given in appendix~\ref{app:RK-subtraction}.
Note that,  in Eq.~\eqref{eq:sch-eq-kappa}, $\bar{\omega}_c$ plays the role of an eigenvalue, while $\bar{E}$ corresponds to the strength of a pure Coulomb interaction. Equivalently, the matrix in $(\kappa,\kappa')$ space multiplying the $\bar{E}$ term can be inverted to treat instead $\bar{E}$ as an eigenvalue.

By applying the  numerical scheme just described to solve the \sch equation for the $s$-states,
we obtain the numerically exact exciton energies and eigenfunctions 
for the Rydberg series of $ns$ states in a TMD monolayer in the presence of a magnetic field of arbitrary value. We have checked that all our results are converged with respect to the number of points on the momentum grid. 
In order to benchmark our approach, we compare our results with the experimental measurements of the exciton diamagnetic shift of Ref.~\cite{Goryca-Crooker_NatCom2019} where the magnetic field dependence of the exciton Rydberg energies for hbN-encapsulated WS$_2$, MoTe$_2$, MoSe$_2$, and MoS$_2$ were obtained for a magnetic field as high as $91$~T. 
As we show, the agreement between our numerics and the experimental data is excellent. In the main text, we illustrate the specific case of an encapsulated WS$_2$ monolayer for $B$ up to $60$~T, while the results for the other TMD monolayers are reported in Appendix~\ref{app:other_exp-diamag}.

The parameters used to describe the hBN-encapsulated WS$_2$ results presented in the main text are listed in Table~\ref{tab:parameters}.
\begin{table}[]
    \centering
    \begin{tabular}{|c|c|c|}
        \hline
                     & hBN-WS$_2$ & GaAs QW \\ \hline \hline
        Effective dielectric constant $\varepsilon$& $4.35$ & $12.9$\\ \hline
        Screening length $r_0$ (nm) & $0.78$ & $0$\\ \hline
        Reduced mass $\mu~(m_0)$ & $0.175$ & $0.041$ \\ \hline
        $1s$ binding energy $\epsilon_b$ (meV) & $178.8$ & $13.5$ \\ \hline
    \end{tabular}
    \caption{Model parameters used in the main text to describe hBN-encapsulated WS$_2$ experiments of Ref.~\cite{Goryca-Crooker_NatCom2019} and typical GaAs quantum well experiments --- see, e.g., Ref.~\cite{Brodbeck_PRL2017}. Here, $m_0$ is the bare electron mass. In the last row we specify the obtained values of the $1s$ exciton binding energies for the two systems --- note that, for QWs, the binding energy $\epsilon_b$ coincides with the exciton Rydberg $R_X$~\eqref{eq:hydrogen_scales}. Note also that we use a different definition of $r_0$ from that of Ref.~\cite{Goryca-Crooker_NatCom2019} ($r_0^{[39]} = r_0 \varepsilon$).}
    \label{tab:parameters}
\end{table}
These parameters have been extracted in Ref.~\cite{Goryca-Crooker_NatCom2019} by fitting the exciton diamagnetic shift with numerically exact solutions of the \sch equation in real space~\eqref{eq:sch-real-space-exciton}. 
We also find that this choice of parameters  listed in Table~\ref{tab:parameters} 
leads to an exceptional agreement of our numerical results with the experimental measurements. 
%
In the same table we also consider, for comparison, the typical parameters describing GaAs quantum well experiments (see, e.g., Ref.~\cite{Brodbeck_PRL2017}).\footnote{\label{footnote3}Note that we consider here the specific case of narrow  GaAs quantum wells as in the experiments of Refs.~\cite{Pietka-Potemski_PRB2017,Brodbeck_PRL2017} and focus on the properties of heavy-hole excitons, which, in this limit, are well described by a parabolic band approximation. Light-hole excitons in these structures are shifted up in energy because of the quantum well confinement and can thus be neglected. Applying this simplified model already demonstrated an excellent agreement with the experiments of Refs.~\cite{Pietka-Potemski_PRB2017,Brodbeck_PRL2017} for the diamagnetic shift of $1s$ and $2s$ heavy-hole excitons up to moderate magnetic fields $B\lesssim 15$~T~\cite{Laird_PRB2022}.} In this case exciton properties are well described by a pure Coulomb interaction potential.
 
\begin{figure}
    \includegraphics[width=\columnwidth]{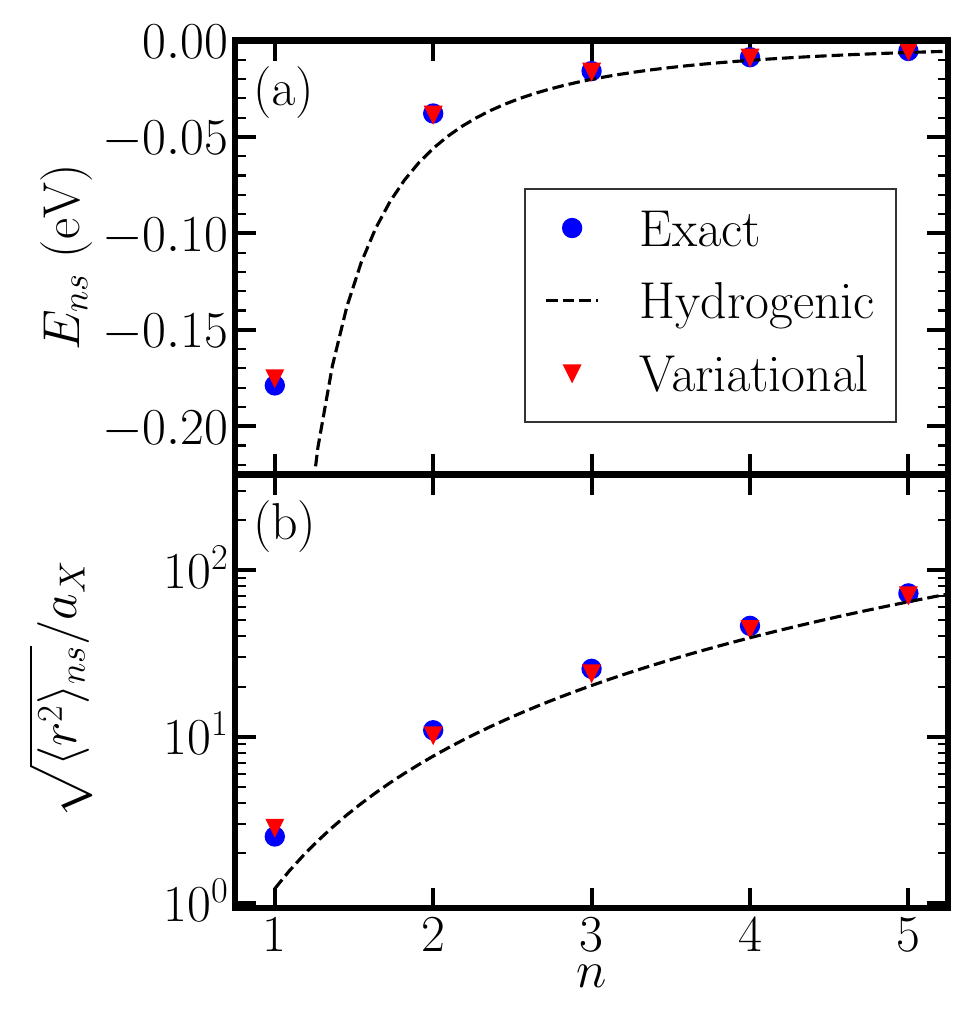}
    \caption{(a) Rydberg exciton energies for $ns$ states at zero magnetic field (blue dots) as a function of the principal quantum number $n$ for system parameters describing the hBN-encapsulated WS$_2$ monolayer experiment of Ref.~\cite{Goryca-Crooker_NatCom2019} (see Table~\ref{tab:parameters}).
    The dashed black line is the hydrogenic energy  for a pure Coulomb interaction and with the same choice of reduced mass $\mu$ and dielectric constant $\varepsilon$; see Eq.~\eqref{eq:hydrogen_energy}. The red triangles are obtained variationally with trial hydrogenic solutions where the exciton Bohr radius is employed as the variational parameter (see appendix~\ref{app:Coulomb}). 
    (b) Exciton root-mean-square radius ($a_X=0.66$~nm). The color code is the same for both panels.}
\label{fig:Rydberg-series}
\end{figure}
\begin{figure*}
    \centering
    \includegraphics[width=\textwidth]{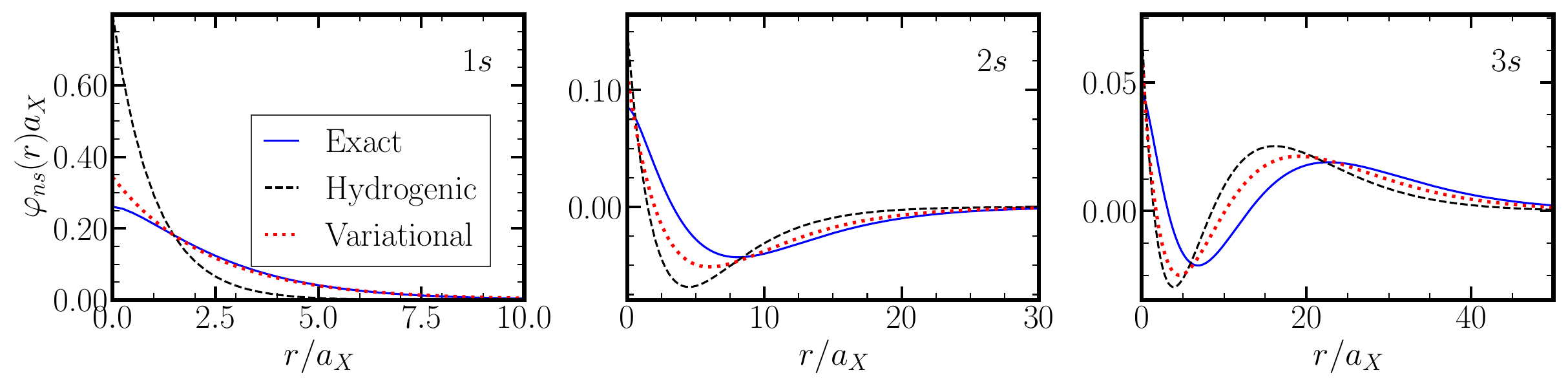}
    \caption{ Electron-hole wavefunctions $\varphi_{ns}(r)$ in real space  at zero magnetic field for the first three states of the Rydberg $ns$ series. Parameters have been fixed to describe  hBN-encapsulated WS$_2$ monolayer (see Table~\ref{tab:parameters}, $a_X=0.66$~nm) and the color code is the same as in Fig.~\ref{fig:Rydberg-series}.}
    \label{fig:exc_wavefunctions}
\end{figure*}
%
\subsubsection{Zero magnetic field}
First we explore the results in the  limit of zero magnetic field and illustrate the deviations of the Rytova-Keldysh interaction potential from the hydrogenic solutions~\cite{Chernikov_PRL2014}, as well as the limitations of a variational approach based on hydrogenic states~\cite{Berkelbach_PRB2013,Zipfel-Chernikov-MF_PRB2018,Shahnazaryan_PRB_2017}. 
We consider the parameters describing the hBN-encapsulated WS$_2$ experiments of Ref.~\cite{Goryca-Crooker_NatCom2019} listed in Table~\ref{tab:parameters}
and we obtain an exciton binding energy of $\epsilon_b = |E_{1s}^{B=0}| = 178.8$~meV for the $1s$ exciton, consistent with the measured value $(180\pm 3)$~meV in Ref.~\cite{Goryca-Crooker_NatCom2019}.
In Fig.~\ref{fig:Rydberg-series}(a) we plot the energies of the first five $s$-wave exciton states $E_{ns}$. To show the deviations from the hydrogenic series obtained for a pure Coulomb interaction (i.e., by setting $r_0=0$), we compare our numerical results for $E_{ns}$ with $E_{ns}^{hyd}$ obtained for the same choice of effective dielectric constant $\varepsilon$ and reduced mass $\mu$; see Eq.~\eqref{eq:hydrogen_energy}. As already discussed in the literature~\cite{Chernikov_PRL2014}, the effect of the non-local screening is clearly visible for the lowest energy states, which are the most affected by  screening. 
For higher excited states, screening effects are reduced as 
they are large compared to  $r_0$. For our particular choice of parameters,  the $n>3$ states are already hydrogen-like. 

In addition, we compare the numerically exact results with those obtained from a 
variational approach with hydrogenic wavefunctions using a variational Bohr radius, as previously employed~\cite{Berkelbach_PRB2013,Zipfel-Chernikov-MF_PRB2018,Shahnazaryan_PRB_2017} 
(see Appendix~\ref{app:Coulomb} for details). The variational parameters obtained with this procedure are summarized in Table~\ref{tab:variational_parameters}.
We observe that the variational results for the exciton energies $E_{ns}$ match very well the numerically exact values, with a maximum deviation for the $1s$ state of about $\sim 2.2\%$. As $n$ grows, the states become more hydrogenic and the matching between exact and variational results improves. The same applies to the exciton root-mean-square radius, plotted in Fig.~\ref{fig:Rydberg-series}(b). Minor deviations occur between the variational and the exact approach, which vanish for excited states, 
in agreement with previous work~\cite{Berkelbach_PRB2013,Chernikov_PRL2014,Zipfel-Chernikov-MF_PRB2018}. 

However, as illustrated in Fig.~\ref{fig:exc_wavefunctions}, the overall shape of the exciton wavefunction is modified by the Rytova-Keldysh potential in a way that cannot be captured by the variational  wavefunction, with differences persisting even for excited states. We will argue in Sec.~\ref{sec:inter_strength} that these differences 
lead to qualitatively different results for the exciton-exciton interaction strength 
when compared with previous results in Ref.~\cite{Shahnazaryan_PRB_2017}.

\subsubsection{Finite magnetic field}
\label{sec:exc-magnetic-field}
\begin{figure}
    \centering
    \includegraphics[width=\columnwidth]{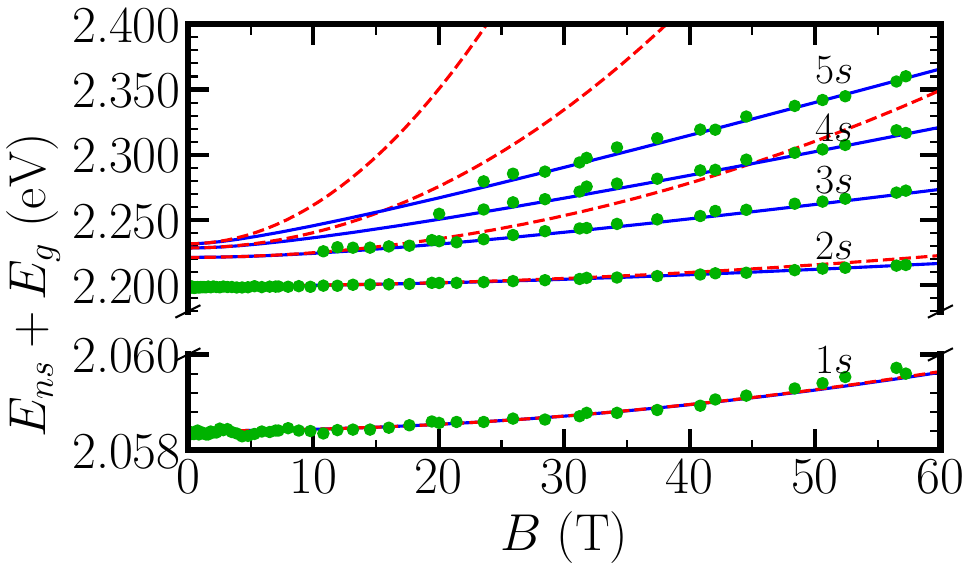}
    \caption{Comparison of the exciton diamagnetic shift for the first five $s$-states obtained from the numerically exact solution of Eq.~\eqref{eq:sch-eq-kappa} (solid blue lines) with the experimental results for hBN-encapsulated WS$_2$ of Ref.~\cite{Goryca-Crooker_NatCom2019} (green dots). The same excellent agreement was also shown in Ref.~\cite{Goryca-Crooker_NatCom2019}, where the \sch equation~\eqref{eq:sch-real-space-exciton} was instead solved in real space. Dashed red lines are results of perturbation theory for a weak magnetic field~\eqref{eq:perturbative_smallB}. The energy gap is set to 
    $E_g=2.23725$~eV. }
    \label{fig:WS2-diamagnetic-shift}
\end{figure}

We now consider the effects of a static magnetic field applied perpendicularly to the TMD monolayer. 
The field strengths are taken to be up to $60$~T as in the experiments of Ref.~\cite{Goryca-Crooker_NatCom2019}.  We plot in Fig.~\ref{fig:WS2-diamagnetic-shift} the exciton energies of the first five $s$-states as a function of the magnetic field,
finding excellent agreement between theory and experiment.
When comparing numerically exact results with the perturbative expansion~\eqref{eq:perturbative_smallB} valid in the low-field limit, we observe that the $1s$ state is accurately described by the quadratic perturbative term. Indeed, the large binding energy of the $1s$ state is such that the condition $\epsilon_b = |E_{1s}^{B=0}|\gg \omega_c$ is satisfied for values of the magnetic field up to $60$~T. 
However, for the $2s$ state, the perturbative expression becomes inadequate for magnetic field strengths larger than $45$~T. For higher energy states, the binding energy $|E_{ns}^{B=0}|$ strongly decreases with $n$ and the perturbative results lose validity for increasingly smaller values of the magnetic field. 

\begin{figure*}
    \centering
    \includegraphics[width=\textwidth]{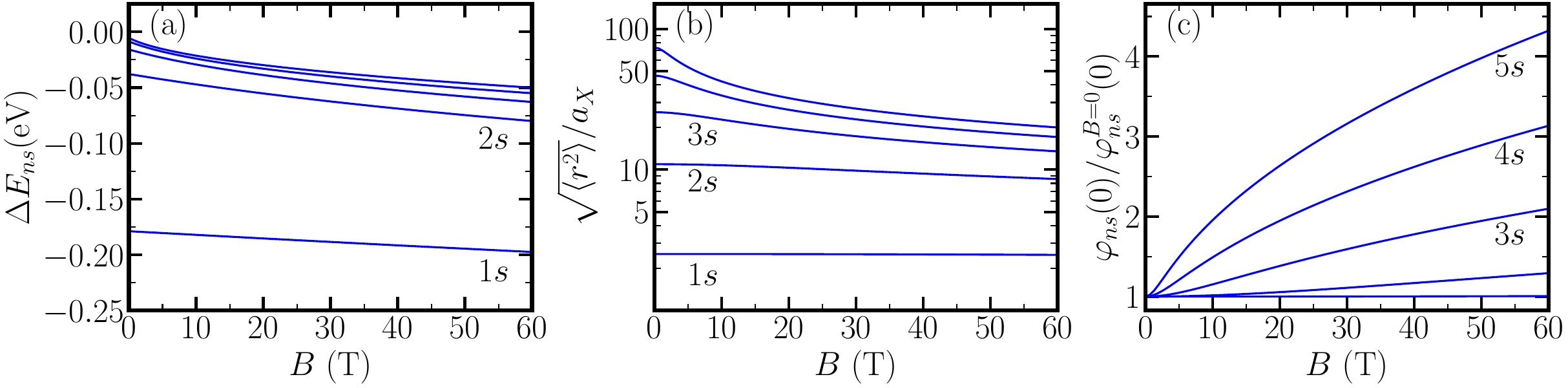}
    \caption{(a) Exciton energy shift from the continuum (the Landau energy of a free electron and hole) $\Delta E_{ns}=E_{ns}-(2n-1)\omega_c$. (b) Root-mean-square electron-hole separation. (c) Normalized exciton oscillator strength. System parameters are for a hBN-encapsulated WS$_2$ monolayer and are listed in Table~\ref{tab:parameters} and $a_X=0.66$~nm.}
    \label{fig:binding-rms-osc-exciton}
\end{figure*}

Even though the exciton energy $E_{ns}$ grows with the magnetic field, the continuum also shifts upwards such that the excitons effectively become more strongly bound with $B$, similarly to the case of pure Coulomb interactions~\cite{Stafford_PRB1990,MacDonald-Ritchie_PRB1986,Laird_PRB2022}.  Indeed, 
for a given $n$ the continuum starts at the Landau energy $E_{Ns}^{\text{Landau}}= \frac{1}{2}(\omega_{c,e}+\omega_{c,h})(N+1)=\omega_c(N+1)$~\cite{landau1991quantum}, where the Landau level index $N$ and the exciton principal quantum number $n$ are related by $N=2 (n-1)$. 
To illustrate this, we plot in Fig.~\ref{fig:binding-rms-osc-exciton}(a) the energy difference $\Delta E_{ns}=E_{ns}-(2n-1)\omega_c$, the absolute value of which coincides with the exciton binding energy. $\Delta E_{ns}$ becomes more negative as the magnetic field increases and excitons become more bound. Note that, for the magnetic field values considered, the $1s$ state energy $E_{1s}$ is well described by the perturbative small field quadratic expression, and thus the dominant magnetic field dependence of  $\Delta E_{1s}=E_{1s}-\omega_c$ at small values of $B$ is linear.

The increase in the exciton binding energy with the magnetic field is accompanied by a reduction in the exciton size, as shown in Fig.~\ref{fig:binding-rms-osc-exciton}(b). This can be evaluated in $\kappa$-space as~\cite{Laird_PRB2022}:
\begin{equation}
\label{eq:exciton-size}
   \frac{\langle r^2 \rangle}{a_X^2}=\frac{1}{a_X^2}\int d\r \,r^2 |\varphi(r)|^2=32\sum_{\bm\kappa}|\bar{\varphi}_{\kappa}|^2\; ,
\end{equation}
where we have used the normalization condition~\eqref{eq:normalization}.

The magnetic field also  affects the exciton oscillator strength, i.e., the exciton's ability to couple to light. This is proportional to the probability of recombination of the electron-hole pair, i.e., to $|\varphi(r=0)|^2$~\cite{klingshirn2012semiconductor}. The probability amplitude $\varphi(r=0)$ can be computed as:
\begin{equation}
    \varphi(r=0)=\frac{1}{a_X}\bar{\varphi}(\rho=0)=\frac{1}{a_X}\sum_{\bm{\kappa}} \bar{\varphi}_{\kappa}\;.
\end{equation}
We show in Fig.~\ref{fig:binding-rms-osc-exciton}(c) the evolution of $\varphi(0)$ with the magnetic field. The increase of the exciton binding energy with the magnetic field is accompanied by an increase of the exciton's ability to couple to light. Similarly to GaAs QWs, the effect of the magnetic field becomes more relevant for excited states that have a smaller binding energy. However, in TMD monolayers, the binding energy and oscillator strength decrease slower with $n$ compared to the hydrogenic series.
Because of this, the effect of an applied magnetic field in modifying the relative wavefunction is weaker than the results found in Ref.~\cite{Laird_PRB2022} for the case of GaAs quantum wells.

Finally, we note that, in general, the diamagnetic shifts of bright exciton states compete with other effects such as electron-hole exchange and the Lamb shift. It is therefore worthwhile to estimate the energy shifts due to such effects and compare them with the diamagnetic shifts that we have already shown in Fig.~\ref{fig:WS2-diamagnetic-shift}. In general, these effects rely on the matrix element for electron-hole transitions and are thus proportional to the exciton oscillator strength~\cite{Combescot2022}, i.e., they are proportional to $|\varphi_{ns}(0)|^2$ for a given $ns$ state.
For the bright 1$s$ exciton, the associated energy shift has been measured to be roughly $\Delta E_{1s}\simeq1$~meV in TMDs~\cite{Ren2023}. While such an energy shift is measurable, it is still much less than the splitting between the $1s$ and $2s$ states in the TMDs (which is larger than $100$~meV), and therefore would only very weakly influence the $1s$ state on the scale of Fig.~\ref{fig:WS2-diamagnetic-shift} (e.g., such a shift could easily be incorporated into a slight variation in the parameters such as $r_0$ without a noticeable effect). Furthermore, for the case of the 1$s$ exciton, the change in oscillator strength from $B=0$ to $B=60$ T is $0.6\%$ [see Fig.~\ref{fig:binding-rms-osc-exciton}(c)], and therefore we would estimate a change in electron-hole exchange and Lamb shift less than 1 $\mu$eV across this range, negligible when compared with the observed diamagnetic shifts. For the excited states, the relative energy shifts are even smaller than for the $1s$ state. For instance, using the hydrogenic expressions~\cite{Parfitt-Portnoi_JMP2002} at $B=0$, we have $E_{ns}\simeq -\frac{R_X}{(2n-1)^2}$, while $\Delta E_{ns}\simeq~1$~meV$\frac{|\varphi_{ns}^{B=0}(0)|^2}{|\varphi_{1s}^{B=0}(0)|^2}\simeq \frac{1~\text{meV}}{(2n-1)^3}$ and thus the relative shift $\frac{\Delta E_{ns}}{E_{ns}}\simeq~\frac{1~\text{meV}}{[R_X (2n-1)]}$ is smaller than 1$\%$, becoming even smaller with increasing $n$. The relative shift furthermore decreases with increasing magnetic field: Even though the electron-hole transition becomes more likely according to Fig.~\ref{fig:binding-rms-osc-exciton}(c), now the corresponding shift should be compared with the cyclotron frequency, which increases linearly with $B$, and consequently we find that the relative shift decreases. Thus, we conclude that it is a reasonable assumption to neglect the Lamb shift and electron-hole exchange.

%
\section{Magnetopolaritons}
\label{sec:polaritons}
We now turn to the effects of a strong perpendicular magnetic field on a TMD monolayer embedded in a microcavity. 
In this scenario, the excitons become coupled to the cavity photon mode. Denoting the strength of the coupling to the $1s$ exciton by $\Omega/2$,
if the Rabi splitting $\Omega$ exceeds both the exciton and photon linewidths, the reversible energy transfer between excitons and cavity photons leads to the formation of magnetopolaritons. 
When the coupling strength becomes a sizable fraction of the exciton binding energy $\epsilon_b$, also referred to as the very strong coupling regime, the light-matter interaction hybridizes different excitonic states, leading to modifications of the electron-hole pair wavefunction~\cite{Levinsen_PRR2019}.
The use of magnetic field as a tool for investigating and manipulating polaritons has already been proposed for III-V heterostructures~\cite{Tignon_PRL1995,Pietka_PRB2015,Pietka-Potemski_PRB2017}. For the same structures, it has also been proposed as a method to verify the regime of very strong light-matter coupling~\cite{Brodbeck_PRL2017,Laird_PRB2022}. Indeed, in this configuration the magnetic field is employed to probe modifications of the electron-hole wavefunction due to the very strong coupling to light. 

The very strong coupling regime can be easily accessed experimentally in III-V and II-VI heterostructures by embedding multiple quantum wells into the microcavity~\cite{Bloch_APL1998,
Saba_Nat2001}, since the Rabi splitting $\Omega$ grows as the square root of the number of quantum wells.  
In particular, in Ref.~\cite{Brodbeck_PRL2017}, it was possible to reach $\Omega/\epsilon_b \simeq 1.3$ (with $\epsilon_b = R_X = 13.5$~meV) by embedding 28 quantum wells in
stacks of four at the seven central antinodes of the cavity light field, thus increasing $\Omega$ from $3.8$~meV for one quantum well, to $\Omega = 17.4$~meV for 28 quantum wells.

In TMD monolayers, the exciton binding energy is about one order of magnitude larger than in III-V 
quantum wells such as GaAs (see Table~\ref{tab:parameters}), and likewise the exciton oscillator strength is also much larger than for QWs. 
Recently, 
an increase of the Rabi splitting has been achieved, from $36$~meV for one WS$_2$ monolayer embedded into a planar microcavity to $72$~meV for 4 embedded monolayers~\cite{
Zhao-Sanvitto_NatComm2023}. If $\epsilon_b = 178.8$~meV, this implies that $\Omega/\epsilon_b \simeq 0.4$. Increasing the number of embedded monolayers, one can reach larger values of $\Omega/\epsilon_b$. We will fix later the specific value $\Omega = 100$~meV ($\Omega/\epsilon_b \simeq 0.6$) and show that clear signatures of the very strong coupling regime and hybridization of different excitonic states are accessible in current experiments~\cite{Zhao-Sanvitto_NatComm2023}.

We now briefly review  
the derivation of the \sch equations for magnetopolaritons. These were first obtained in Ref.~\cite{Laird_PRB2022}. Here, however, we employ a second quantization formalism in  real space~\eqref{eq:Hamil_all_terms} since this will allow us to generalize our results to the case of two magnetopolaritons and to evaluate their interaction strength. A polariton state with $\vect{K}=0$, 
\begin{equation}\label{eq:pol-state}
    |P_{\vect{0}}^{} \rangle = \hat{P}_{\0}^\dag \ket{0}\; ,
\end{equation}
can be written by considering a superposition between the exciton state~\eqref{eq:exciton-state} and a photon:
\begin{multline}
\label{eq:pol-creation-op}  
    \hat{P}_{\0}^\dag = \frac{1}{\sqrt{\area}}\int d\r_e d\r_h e^{i\left[\frac{e}{2c}\mathbf{B}\times(\r_e-\r_h)\right]\cdot \frac{m_e\r_e+m_h\r_h}{m_e+m_h}}  \\
    \times  \varphi(|\mathbf{r}_e-\mathbf{r}_h|)\hat{\Psi}_e^{\dag}(\mathbf{r}_e)\hat{\Psi}_h^{\dag}(\mathbf{r}_h)
    + \gamma \hat{a}^{\dag}\vphantom{\frac{1}{\sqrt{\area}}} \; ,
\end{multline}
where $\varphi(r)$ is the wavefunction describing the electron and hole relative motion and $\gamma$ is the photon amplitude. The normalization requires that
\begin{equation}
\label{eq:normalization-polaritons}
    \bra{P_{\vect{0}}^{}} \ket{P_{\vect{0}}^{}} = 1  = \int d\r |\varphi (r)|^2+|\gamma|^2\; .
\end{equation}

The coupled equations for the exciton and photon amplitudes can be found by 
minimizing the functional $\bra{P_{\vect{0}}^{}}E - \hat{H} \ket{P_{\vect{0}}^{}}$, obtaining:
\begin{subequations}
\label{eq:coupledEqs-RealSpace-Polariton}
\begin{align}
    &\left[E+\frac{1}{2\mu}\left(\frac{d}{dr^2}+\frac{1}{r}\frac{d}{dr}\right)-\frac{\mu \omega_c^2}{2}r^2-V(r)\right]\varphi(r) \notag\\&=g\gamma \delta(\r)\label{eq:first-coupledEQ-pol} \\
    &(E-\omega)\gamma=g\int d\r \varphi(r)\delta(\r) \label{eq:second-coupledEQ-pol}\; .
\end{align}
\end{subequations}
As already discussed in Ref.~\cite{Laird_PRB2022}, 
the presence of the Dirac delta $\delta(\r)$ in Eq.~\eqref{eq:first-coupledEQ-pol} implies that the relative wavefunction $\varphi(r)$ diverges as $\ln(r)$ when $r\rightarrow0$. The divergent part of $\varphi(r)$ can be isolated by considering instead 
\begin{equation}
\label{eq:ansatz-renorm-phi}
    \varphi(r)=\beta(r)-\frac{g\gamma \mu}{\pi} K_0\left(\frac{r\sqrt{\bar{\epsilon}_{b}}}{a_X}\right)\; ,    
\end{equation}
where $K_0(x)$ is the zeroth-order modified Bessel function of the second kind. Here, $K_0(r)$ diverges as $\ln(r)$ when $r$ goes to zero, thus canceling out the delta function $\delta(\r)$ term in Eq.~\eqref{eq:first-coupledEQ-pol}. On the other hand, 
$\beta(r)$ is a regular function at $r=0$.

The logarithmic divergence of $K_0(r)$ when $r\rightarrow 0$ leads to the formal divergence of the integral on the right-hand side of Eq.~\eqref{eq:second-coupledEQ-pol}, which requires us to renormalize the photon energy $\omega$, as already discussed in Ref.~\cite{Levinsen_PRR2019}. 
Note that because the divergence occurs at zero distance, the magnetic field does not interfere with the renormalization process.
Indeed, following~\cite{Laird_PRB2022}, the formal divergence of Eq.~\eqref{eq:second-coupledEQ-pol} can be dealt with by properly relating the physical observables of the system with the parameters of the microscopic model. Observables in experiments  are the photon energy in the presence of an active medium, or, equivalently, the photon energy detuning $\delta$ measured from the $1s$ exciton energy $E_{1s}^{B=0} = -\epsilon_b$, and the Rabi splitting $\Omega$. Both quantities are measured in the absence of a magnetic field. In the limit $\Omega \ll \epsilon_b$, i.e., away from the very strong coupling regime, one expects to recover the polariton energies obtained  from a two coupled oscillator model (2-COM)
\begin{equation}
\label{eq:2-COM}
  E\begin{pmatrix}
        \beta \\
        \gamma
    \end{pmatrix}= \begin{pmatrix}
        -\epsilon_b &\Omega/2 \\ \Omega/2 & \delta -\epsilon_b
    \end{pmatrix}\begin{pmatrix}
        \beta \\
        \gamma
    \end{pmatrix}\; ,
\end{equation}
where $|\gamma|^2$ and $|\beta|^2=1-|\gamma|^2$ are the photon and exciton photon fractions, respectively. Indeed, in this limit, and away from the resonance with excited exciton states, the exciton wavefunction is barely modified by light.
The eigenvalues of the 2-COM are the usual lower polariton (LP) and upper polariton (UP) modes with energies:
\begin{equation}
\label{eq:LP-UP_2-COM}
    E_{LP,UP}=-\epsilon_b + \frac{1}{2}(\delta\mp\sqrt{\delta^2+\Omega^2}) \; ,
\end{equation}
and Hopfield coefficients:
\begin{equation}
\label{eq:Hope}
    \gamma_{LP,UP}^{}=\mp \sqrt{\frac{1}{2}\left(1\mp\frac{\delta^2}{\sqrt{\delta^2+\Omega^2}}\right)}\; .
\end{equation}
One can show~\cite{Levinsen_PRR2019} that Eqs.~\eqref{eq:coupledEqs-RealSpace-Polariton} at zero magnetic field and in the limit $\Omega \ll \epsilon_b$ recover those of the  2-COM
for the following expressions of the 
 detuning $\delta$ and Rabi splitting:
\begin{subequations}
\label{eq:renorm}
\begin{align}
\label{eq:def-delta-real-space}
    \delta &=\omega - \frac{g^2\mu}{\pi}\int d\r \ K_0\left(\frac{r\sqrt{\epsilon_b/R_X}}{a_X}\right)\delta(\r)\ + \epsilon_b\\
\label{eq:def-Rabi}    
    \Omega &=2 g \varphi_{1s}^{B=0}(0)\; ,
\end{align}  
\end{subequations}
where $\varphi_{1s}^{B=0}(0)$ is the exciton wavefunction at zero separation in absence of magnetic field and light-matter coupling. 
The integral in Eq.~\eqref{eq:def-delta-real-space} is formally divergent and it exactly cancels out the divergence on the right-hand side of  Eq.~\eqref{eq:second-coupledEQ-pol} when the bare photon energy $\omega$ is written in terms of the renormalized photon detuning $\delta$. Instead, Eq.~\eqref{eq:def-Rabi} does not diverge, as $\varphi_{1s}^{B=0}(r)$ has a well defined limit for $r\to 0$. 

Introducing these definitions into Eq.~\eqref{eq:coupledEqs-RealSpace-Polariton}, we can now carry out the same change of variable $\r \mapsto \vegr{\rho} \mapsto \vegr{\kappa}$ considered in Sec.~\ref{sec:excitons} in absence of strong light-matter coupling, so that the coupled equations to solve are: 
\begin{subequations}
\label{eq:kappa-space-polaritons}
\begin{align}
      &4 \pi \bar{E}\sum_{\bm{\kappa}'} \frac{\bar{\varphi}_{\kappa'}}{|\bm{\kappa}-\bm{\kappa}'|}-\kappa^2\bar{\varphi}_{\kappa}-4 \omega_c^2 \bar{\varphi}_k-\sum_{\bm{\kappa}'} \tilde{V}_{|\bm{\kappa}-\bm{\kappa}'|} \bar{\varphi}_{\kappa'} \notag\\& =\frac{\bar{\Omega}}{4\bar{\varphi}_{1s}^{B=0}(0)}\gamma \\
      &\left[\bar{E}-\bar{\delta}-\left(\frac{\bar{\Omega}}{2\sqrt{2} \varphi_{1s}^{B=0}(0)}\right)^2\sum_{\bm{\kappa}}F(\kappa;\bar{\epsilon}_b)+\bar{\epsilon}_b\right]\gamma \notag\\& =\frac{\bar{\Omega}}{2  \bar{\varphi}_{1s}^{B=0}(0)}\sum_{\bm{\kappa}} \bar{\varphi}_{\kappa}\;, 
\end{align}
\end{subequations}
where, as before we rescale quantities by the Rydberg energy and Bohr radius~\eqref{eq:hydrogen_scales}, so that $\bar{\varphi}_{1s}^{B=0} (0) = a_X\varphi_{1s}^{B=0} (0)$, $\bar{\Omega}=\Omega/R_X$, $\bar{\delta}=\delta/R_X$, and $\bar{\epsilon}_b = \epsilon_b/R_X$, and where
\begin{equation}
     F(\kappa;\bar{\epsilon}_b)=\frac{1}{\kappa^2}+\frac{\bar{\epsilon}_b\pi}{\kappa^3}\left[Y_0\left( \frac{2\bar{\epsilon}_b}{\kappa}\right)  - H_0\left(\frac{2\bar{\epsilon}_b}{\kappa}\right)  \right]\; .   
\end{equation}

As we did for the exciton case, we can numerically solve the coupled equations~\eqref{eq:kappa-space-polaritons} by diagonalization once we have discretized the rescaled momentum $\kappa$ on a grid and applied the subtraction trick described in appendix~\ref{app:RK-subtraction}. 
The normalization~\eqref{eq:normalization-polaritons} can be implemented by the rescaling
\begin{align}
&\bar{\varphi}_{\kappa}\rightarrow\bar{\varphi}_{\kappa}/\mathcal{N} &\gamma\rightarrow\gamma/\mathcal{N},
\end{align}
where $\mathcal{N}^2=8 \pi \sum_{\kappa,\kappa'}\frac{\bar{\varphi}_{\kappa} \bar{\varphi}_{\kappa'}^{*}}{|\bm{\kappa}-\bm{\kappa}'|}+|\gamma|^2$. 
As for the exciton problem, the polariton energy $E$ can be converted back into an eigenvalue by matrix inversion.

\begin{figure}
    \centering
    \includegraphics[width=\columnwidth]{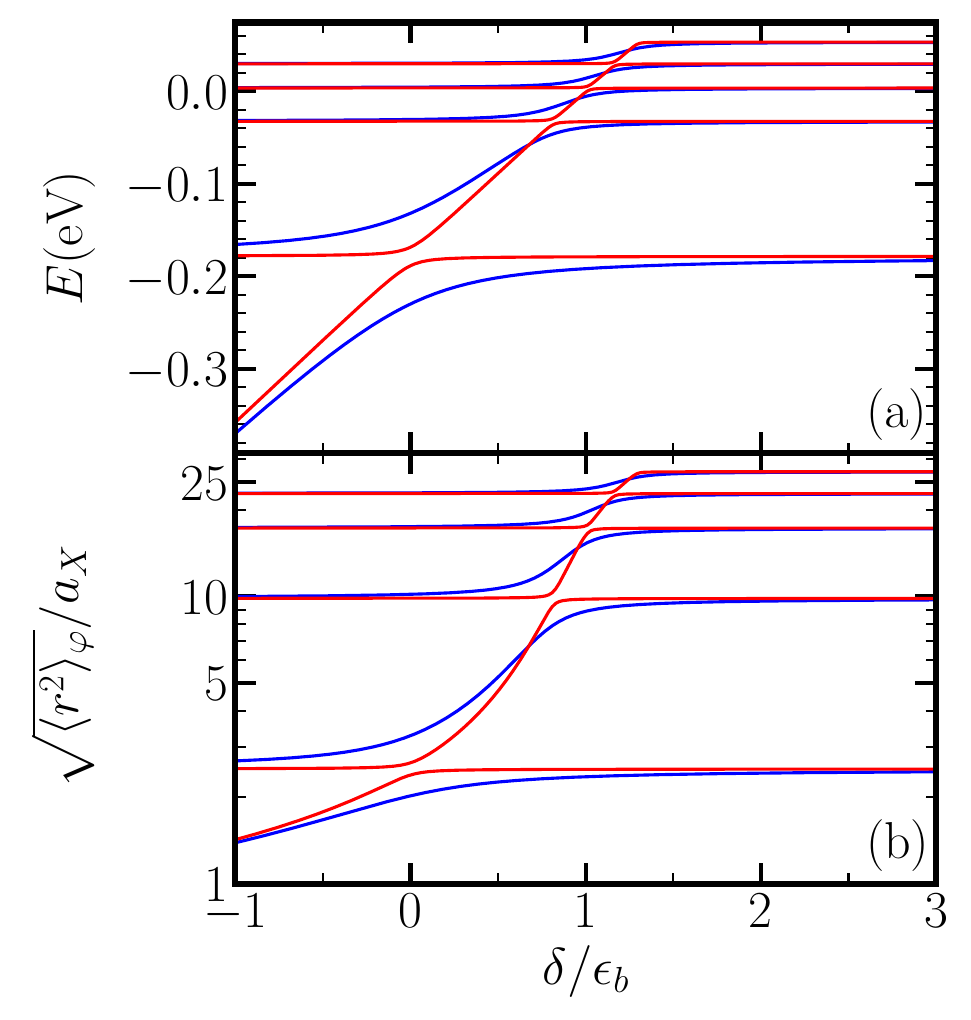}
    \caption{(a) Polariton energies and (b) average electron-hole separation as a function of the exciton-photon detuning for the first five polariton states  at fixed magnetic field $B=30$~T and Rabi coupling $\Omega=20$~meV (red) and $\Omega=100$~meV (blue). From bottom to top, the lines in the two panels correspond to the same states. System parameters for hBN-encapsulated WS$_2$ monolayer are listed in Table~\ref{tab:parameters} and $a_X=0.66$~nm.}
    \label{fig:EN-RMS-polariton}
\end{figure}

In Fig.\ref{fig:EN-RMS-polariton}(a) we show the polariton energy spectrum at a fixed magnetic field as a function of the exciton-photon detuning for two different values of the Rabi splitting $\Omega$. The polariton energies display the expected anticrossing behavior, with the LP interpolating between the photon and the $1s$ exciton state, while the excited UP states interpolate between the $ns$ and the $(n+1)s$ exciton states. The hybridization between different exciton states is evident when the Rabi splitting becomes comparable with the exciton binding energy.

Similarly, to quantify the changes in the matter component of polaritons induced by coupling to light, we examine in Fig.~\ref{fig:EN-RMS-polariton}(b) the detuning dependence of the corresponding electron-hole root-mean-square separation. This is defined as
\begin{equation}
\label{eq:radius_matter_part}
    \frac{\langle r^2 \rangle_{\varphi}}{a_X^2}\equiv\int d\r\frac{r^2 |\varphi(r)|^2}{a_X^2(1-|\gamma|^2)}=\sum_{\bm{\kappa}}\frac{32 |\bar{\varphi}_{\kappa}|^2}{1-|\gamma|^2}\; ,
\end{equation}
where the term $1-|\gamma|^2$ in the denominator comes from the normalization condition~\eqref{eq:normalization-polaritons}. 
Similarly to what observed in III-V heterostructures~\cite{Laird_PRB2022}, we find that the electron-hole separation in the LP is always smaller than that of the $1s$ uncoupled exciton size, with a deviation that increases with negative detunings when the LP becomes more photonic-like. At positive detuning, deviations from the $1s$ size increase for larger $\Omega$. For excited polariton states, the electron-hole separation interpolates as a function of detuning between the $ns$ and the $(n+1)s$ uncoupled exciton sizes, with a typical avoided crossing behavior with larger splitting 
when the Rabi coupling $\Omega$ 
increases.

\begin{figure}[t]
    \centering
    \includegraphics[width=\columnwidth]{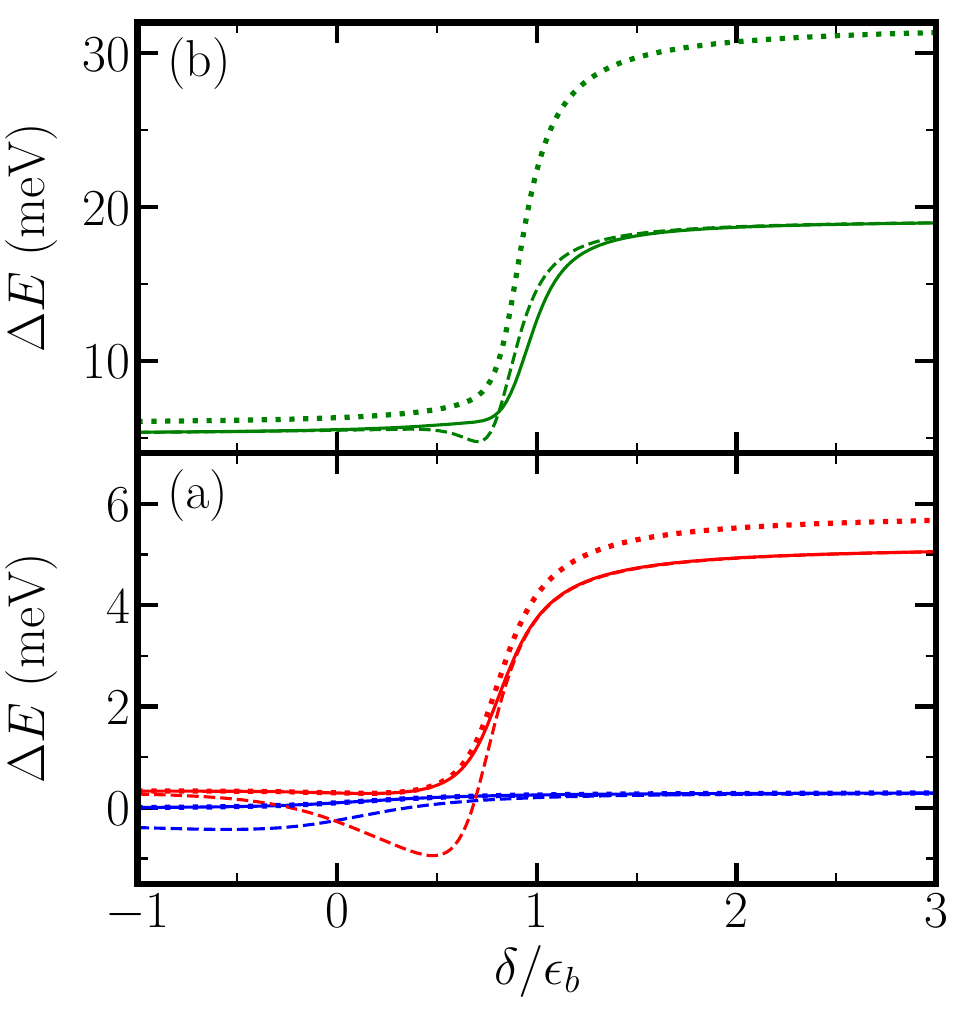}
    \caption{Diamagnetic shift for the first 3 polariton states as a function of detuning at $B=30$~T and Rabi coupling $\Omega=100$~meV: LP  (blue), first excited UP (red) are plotted in panel (a), while the second excited UP state is shown in panel (b). Solid lines are the numerically exact results, dashed lines are those obtained with a 4-COM~\eqref{eq:4-COM}, i.e., perturbative results in the coupling to light, and dotted lines correspond to perturbation theory in the magnetic field~\eqref{eq:diamag-shift-pol}. System parameters for hBN-encapsulated WS$_2$ monolayer are  listed in Table~\ref{tab:parameters}. }
    \label{fig:Pol-diamag-shift}
\end{figure}
We analyze the behavior of the polariton diamagnetic shift in Fig.~\ref{fig:Pol-diamag-shift}. 
Our numerically exact results show that, as a function of detuning, the diamagnetic shifts of the different polariton branches interpolate near monotonically between the diamagnetic shifts of two adjacent excitonic states --- in the case of the LP, the interpolation is between zero, as the photon mode is insensitive to the magnetic field and the diamagnetic shift of the $1s$ excitonic state. Note that, for the parameter choice of Fig.~\ref{fig:Pol-diamag-shift}a, the diamagnetic shift of the LP branch is in the $\mu$eV scale and thus almost imperceptible in the meV scale of Fig.~\ref{fig:Pol-diamag-shift}a. 
For the first excited UP state there is a very small non-monotonicity of the diamagnetic shift, which is progressively lifted with increasing values of $\Omega$.
This behavior originates from two competing effects, namely the non-monotonic change of the photon fraction and the increase of the electron-hole distance with increasing detuning.

It is instructive to compare our 
numerical
results with two perturbative approaches, one treating the magnetic field perturbatively and the other neglecting the light-induced modifications of the exciton wavefunction. In the first case one can use the following perturbative expression for the polariton diamagnetic shift~\cite{Laird_PRB2022}: 
\begin{equation}
\label{eq:diamag-shift-pol}
    \Delta E\simeq\frac{1}{2}\mu \omega_c^2 \langle r^2\rangle_{pol}=\frac{1}{2}\mu \omega_c^2 (1-|\gamma|^2)\langle r^2\rangle_{\varphi}\; ,
\end{equation}
where $\langle r^2 \rangle_{\varphi} $ 
is defined in Eq.~\eqref{eq:radius_matter_part}. This approximation is expected to work well only when light mixes with those excitonic states well described by the small $B$ expansion~\eqref{eq:perturbative_smallB}, i.e., for our choice of parameters, the  $1s$ exciton state only. Note that it captures the qualitative behavior of the diamagnetic shift with the detuning at a fixed magnetic field, but it can give quantitatively wrong results.
Thus, as shown in Fig.~\ref{fig:Pol-diamag-shift}a, it describes the LP state for all values of detunings and the first excited UP state for detunings $\delta < \epsilon_b$, 
while it fails at larger detunings 
as light couples additionally with the $2s$ exciton state. 
Naturally, for the second excited UP polariton state shown in Fig.~\ref{fig:Pol-diamag-shift}b, this perturbative approach fails 
in the whole range of detuning. 

The second perturbative approximation employed involves neglecting the light-induced modifications of the exciton wavefunction. This approximation is expected to work well only when $\Omega \ll \epsilon_b$, while it provides qualitatively incorrect results for the diamagnetic shift and, in some cases, physically inaccurate outcomes~\cite{Laird_PRB2022} when $\Omega$ is comparable to 
$\epsilon_b$. 
In this perturbative approach, we calculate the energies $E_{ns}$ and oscillator strengths $\varphi_{ns} (0)/\varphi_{ns}^{B=0} (0)$ of the exciton states in the absence of light-matter coupling using the methods described in Sec.~\ref{sec:excitons}. Then, if we want to describe the first $n$ polariton states, we include these parameters in an   $(n+1)$-coupled oscillator model (COM) . For example, in order to describe the first three polariton states, we can consider a $4$-COM:
\begin{equation}
\label{eq:4-COM}
    \begin{pmatrix}
        E_{1s} & 0  & 0 & \Omega_{1s}/2\\
        0 &  E_{2s} & 0 & \Omega_{2s}/2\\
        0 & 0 &  E_{3s} & \Omega_{3s}/2 \\
        \Omega_{1s}/2 & \Omega_{2s}/2 & \Omega_{3s}/2 & \delta -\epsilon_b\\
    \end{pmatrix}\; ,
\end{equation}
where $\Omega_{ns} = \Omega \varphi_{ns} (0)/\varphi_{ns}^{B=0} (0)$. From the eigenvalues of this matrix one can evaluate the perturbative diamagnetic shifts shown in Fig.~\ref{fig:Pol-diamag-shift}. The comparison with the exact results illustrates that the 4-COM fails to describe qualitatively the diamagnetic shift and, for the LP and first excited UP, even leads to an unphysical negative diamagnetic shift.
For the first and second excited UP polariton states the diamagnetic shift grows non-monotonically due to the competition between the exciton fraction $1-|\gamma|^2$ and the average electron-hole separation $\langle r^2 \rangle_{\varphi}$; see Eq.~\eqref{eq:diamag-shift-pol}. 
The non-monotonic behavior is progressively lifted for increasing values of $\Omega$, since it increases the average electron-hole separation of the excited UP states; see Fig.~\ref{fig:EN-RMS-polariton}(b). The COM is not able to recover this behavior since it does not consider the back action of light on matter and predicts instead an unphysical negative detuning. This problem cannot be solved by increasing the number of coupled oscillators~\cite{Laird_PRB2022}.
Comparing our numerical results with perturbative approaches in either light-matter coupling or magnetic field effects highlights the need to go beyond these methods and employ our microscopic and exact theory.

\section{Exciton interaction strength}
\label{sec:X-inter_strength}
We now turn to the interaction properties of magnetoexcitons  in TMD monolayers and in III-V and II-VI QWs. 
To this end, we employ the Born approximation~\cite{Ciuti_PRB1998,Tassone-Yamamoto_PRB1999}, which is the state of the art when one considers the complex interplay between the underlying electronic degrees of freedom in the problem. Crucially, the Born approximation provides an upper limit on the exciton-exciton interaction strength for identical ground-state excitons where no biexciton bound state exists~\cite{Li-Bleu-Levinsen-Parish_PRB2021}. Alternatively, one can instead consider model exciton-exciton potentials, which generally give results that are of comparable magnitude~\cite{Schindler2008}. 
Our calculations  extend the results of Ref.~\cite{Ciuti_PRB1998,Tassone-Yamamoto_PRB1999} to TMD monolayers, where interactions are described by the Rytova-Keldysh potential, and to a finite magnetic field. 
Importantly, we find that for TMD monolayer excitons at zero magnetic field, the use of hydrogenic variational wavefunctions~\cite{Shahnazaryan_PRB_2017}  gives qualitatively wrong results for the exciton-exciton interaction dependence on the screening length $r_0$. 

Within the Born approximation, one approximates the exact two-exciton state by that of uncorrelated excitons 
\begin{equation}
    \ket{X_{\vect{0}}^{2}} = \hat{X}_\0^{\dag}\hat{X}_\0^{\dag} \ket{0} \; ,
\end{equation}
where the creation operator of a zero momentum exciton $\hat{X}_\0^{\dag}$ is defined in Eq.~\eqref{eq:Xoperator}. 
This state has a normalization given by
\begin{widetext}
\begin{equation}
\label{eq:normal_2X_state}
    \bra{X_{\vect{0}}^{2}} \ket{X_{\vect{0}}^{2}} =2-\frac{2}{\area^2}\int d \r_e d\r_h d\r_e'd\r_h'e^{-i[\frac{e}{2c}\bm{B}\times(\r_e-\r_e')]\cdot(\r_h-\r_h')} \varphi^{*}(|\r_e'-\r_h'|)\varphi^{*}(|\r_e-\r_h|)\varphi(|\r_e'-\r_h|)\varphi(|\r_e-\r_h'|)\; .
\end{equation}
\end{widetext}
Here, $\varphi(r)$ is the exciton wavefunction describing the electron-hole relative motion and satisfying the \sch equation~\eqref{eq:sch-real-space-exciton}; in practice, $\varphi(r)$ is determined by solving the \sch equation~\eqref{eq:sch-eq-kappa} in the rescaled reciprocal space $\kappa$, and then transforming back $\kappa \mapsto \rho \mapsto r$ to real space. 
Note that the second term in the above normalization arises from the fact that the operator $\hat{X}_{\0}^{}$ only approximately satisfies bosonic commutation relations due to its composite nature. Note also that the normalization~\eqref{eq:normal_2X_state} has two terms. The first scales as $O(\area^{0})$ in the area, while the second is $O(\area^{-1})$, since one can show that only 3 of the 4 real space integrals are independent.

As in previous work~\cite{Ciuti_PRB1998,Tassone-Yamamoto_PRB1999,Combescot_EurPhys2007,Levinsen_PRR2019}, we perform our calculations by evaluating the energy of two excitons as twice the energy of a single exciton $E$ plus the interaction energy between two excitons:
\begin{equation}
\label{eq:Born-approx-gXX}
     \frac{\langle X_{\vect{0}}^{2}|\hat{H}_m|X_{\vect{0}}^{2}\rangle}{\langle X_{\vect{0}}^{2}|X_{\vect{0}}^{2}\rangle}=2E+\frac{g_{XX}^{}}{\area}\; .
\end{equation}
The last term comes from the interaction energy $g_{XX}^{} N(N-1)/2\area$ of $N=2$ identical bosons which scales as the inverse system area $\area$. Note that we are neglecting the momentum transfer between excitons. Since the direct interaction vanishes at zero transferred momentum~\cite{Erkensten_PRB2021}, this implies that we are only including exchange interactions. This is reasonable since the region of small momenta is the most relevant for the optical excitation, where the exchange interaction dominates~\cite{Ciuti_PRB1998,Tassone-Yamamoto_PRB1999,Erkensten_PRB2021}. 
We can extract the exciton-exciton interaction strength from Eq.~\eqref{eq:Born-approx-gXX}.
Because the second term of the normalization $\bra{X_{\vect{0}}^{2}} \ket{X_{\vect{0}}^{2}}$~\eqref{eq:normal_2X_state} is of order $O(\area^{-1})$, as first observed in Ref.~\cite{Levinsen_PRR2019}, we arrive at the following approximated expression of the exciton-exciton interaction strength:
\begin{equation}
\label{eq:ex-ex-general-exp}
    \frac{g_{XX}^{}}{\area}\simeq \Frac{1}{2} \langle X_{\vect{0}}^{2}|\hat{H}_m|X_{\vect{0}}^{2}\rangle - E \langle X_{\vect{0}}^{2}|X_{\vect{0}}^{2}\rangle\; .
\end{equation}
One can show that the terms of $O(1)$ on the r.h.s. of this expression cancel with each other, leaving only the correct contribution $O(\area^{-1})$. We obtain the following final expression for the exciton-exciton interaction strength:
\begin{widetext}
\begin{multline}
\label{eq:gXX}
   \frac{g_{XX}^{}}{\area}=-\frac{1}{\mathcal{A}^2}\int d\r_e d \r_h d\r_e'd\r_h'e^{-i\frac{e}{2c}[\mathbf{B}\times(\r_e'-\r_e)]\cdot(\r_h'-\r_h)} 
      \left[V(|\r_e-\r_h|)+V(|\r_e'-\r_h'|)-V(|\r_e-\r_e'|)-V(|\r_h-\r_h'|)\right] \\
      \times \varphi^{*}(|\r_e-\r_h|)\varphi^{*}(|\r_e'-\r_h'|) \varphi(|\r_e'-\r_h|)\varphi(|\r_e-\r_h'|)\; .
\end{multline}
\end{widetext}
Using the appropriate change of variables, it can be easily seen that the integrand of Eq.~\eqref{eq:gXX} depends only on three integration variables, and that this term scales as inverse area, as it should. Further, one can show that Eq.~\eqref{eq:gXX} is real.
Equation~\eqref{eq:gXX} recovers known results in the limit of zero magnetic field. 
In particular, $g_{XX}^{B=0}$ coincides with  the electron-exchange contribution to the exciton interaction obtained in Ref.~\cite{Ciuti_PRB1998}, as the direct and exciton-exchange terms are zero for zero transferred momentum. Further, as discussed in the next section, when this expression is rewritten in momentum space, it coincides with the one derived in Ref.~\cite{Tassone-Yamamoto_PRB1999}.

In the following, we divide the discussion of results by considering first the case of zero magnetic field in Sec.~\ref{sec:gXX-zeroMF} and then finite magnetic field in Sec.~\ref{sec:gXX-finiteMF}. We will compare the results obtained for TMD monolayers with those  obtained for GaAs quantum wells, for which one can employ a pure Coulomb interaction.
Note that we consider exclusively the interaction properties of the  lowest energy exciton state, i.e., the $1s$ state. This is because the Born approximation provides an upper bound to the interaction strength for this state~\cite{Li-Bleu-Levinsen-Parish_PRB2021}, while being an uncontrolled approximation for excited states.

\subsection{Zero magnetic field}
\label{sec:gXX-zeroMF}
In the absence of a magnetic field, it is convenient to rewrite Eq.~\eqref{eq:gXX} in momentum space, 
\begin{multline}
\label{eq:gXX-zeroB}
    \frac{g_{XX}^{B=0}}{\area} 
    = \frac{1}{\area^3} \sum_{\k,\k'} V_{|\k-\k'|} |\varphi_{k}^{}|^2|\varphi_{k'}^{}|^2\\
    - \frac{1}{\area^3} \sum_{\k,\k'} V_{|\k-\k'|} \varphi_{k}^{*} \varphi_{k'}^{*} \varphi_{k'}^2 \; ,
\end{multline}
where $V_k=\int d\r e^{-i\k\cdot\r} V(r)$ and $\varphi_{k}=\int d\r e^{-i\k\cdot\r}\varphi(r)$ are the Fourier transforms from real $\r$ to momentum space $\k$.
As already shown in Ref.~\cite{Levinsen_PRR2019}, this expression coincides with the expression derived in Ref.~\cite{Tassone-Yamamoto_PRB1999}.
As expected, when $r_0 \to 0$, one recovers the result 
for a pure Coulomb potential~\cite{Tassone-Yamamoto_PRB1999}:
\begin{equation}
\label{eq:gXX-hyd}
    \frac{g_{XX}^{hyd}}{\area}=\frac{6.0566}{2\mu}  \; .
\end{equation}
This result is universal, in the sense that it depends only on the exciton reduced mass, but not on the exciton Rydberg energy~\eqref{eq:hydrogen_scales}.

\begin{figure}
    \centering
    \includegraphics[width=\columnwidth]{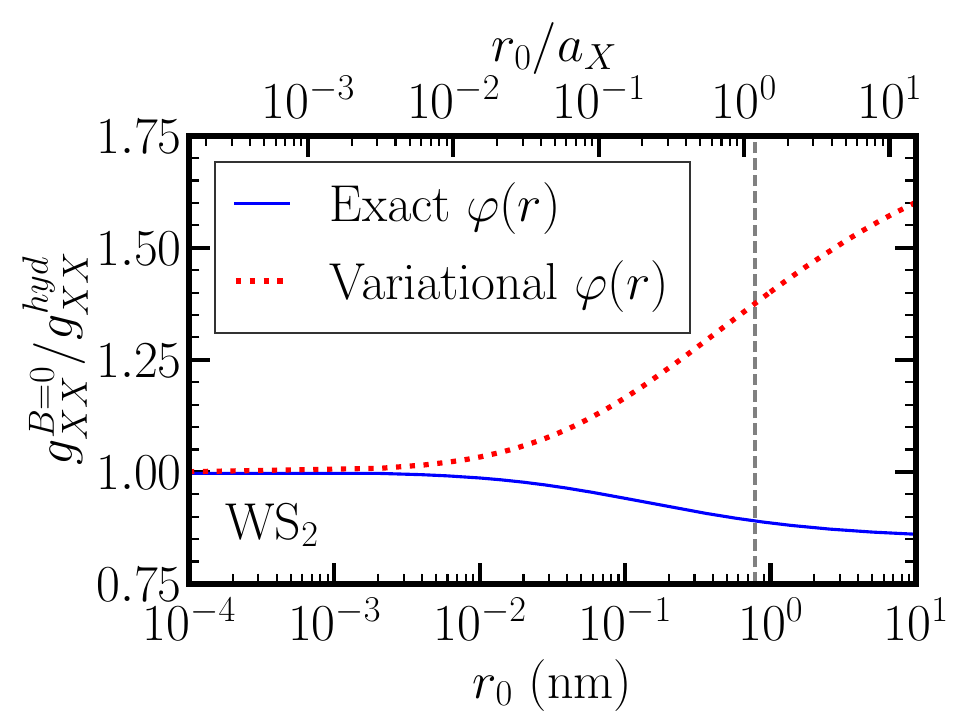}
    \caption{Screening length dependence of the exciton-exciton interaction strength for hBN-encapsulated WS$_2$ at zero magnetic field. The interaction strength is calculated within the Born approximation and is normalized by $g_{XX}^{hyd}$~\eqref{eq:gXX-hyd}, i.e., the corresponding exciton-exciton interaction strength for pure Coulomb interaction.
    Our numerically exact evaluation of Eq.~\eqref{eq:gXX-zeroB} (blue solid line) is compared with that obtained with a variational calculation (red dotted line) as in Ref.~\cite{Shahnazaryan_PRB_2017}. 
    The vertical dashed line indicates the value of $r_0$ describing the hBN-encapsulated WS$_2$ monolayer, see Table~\ref{tab:parameters} and $a_X=0.66$~nm. The scale of the upper $x$-axis is the rescaled screening length.}
    \label{fig:gxx_evol_r0}
\end{figure}

When we consider instead the case of the Rytova-Keldysh potential, we find that the effect of the screening length $r_0$ is to reduce the  value of $g_{XX}^{B=0}$, which  decreases monotonically as a function of $r_0$, see Fig.~\ref{fig:gxx_evol_r0}. Note that $g_{XX}^{B=0}/g_{XX}^{hyd}$ depends universally on the rescaled screening length $r_0/a_X$, remaining independent of the specific parameters chosen. For the specific case of hBN-encapsulated WS$_2$ considered throughout this work, we obtain an exciton interaction strength 
\begin{equation}
\label{eq:WS2-gXX_Bzero}
    \frac{\left. g_{XX}^{B=0}\right|_{\text{WS}_2}}{\area} = \frac{5.39}{2\mu} \; .
\end{equation}
This result contradicts that obtained from a variational approach based on the hydrogenic exciton state with a trial Bohr radius, which instead predicted an increase of the interaction strength with increasing $r_0$~\cite{Shahnazaryan_PRB_2017}.  

The qualitatively different results reveal the sensitivity of the integrals in Eq.~\eqref{eq:gXX-zeroB} to the exact shape of the exciton wavefunction, as shown in Fig.~\ref{fig:exc_wavefunctions},  while the exciton energy and size are largely insensitive to the precise wavefunction, see Fig.~\ref{fig:Rydberg-series}. 
Reference~\cite{Shahnazaryan_PRB_2017} explained their calculated 
increase of $g_{XX}^{B=0}$ 
as due to the characteristic non-local screening of the Rytova-Keldysh potential, which leads to the spreading of the electron-hole  wavefunction in real space. 
From Fig.~\ref{fig:exc_wavefunctions}, it is evident that such a spreading does occur, although it is also clear that the exact exciton wavefunction does not exhibit a hydrogenic shape.  The difference in shape is particularly pronounced near $r=0$, where the interaction potential diverges. We note that the spreading of the wavefunction increases the absolute value of each term in the expression~\eqref{eq:gXX-zeroB}, however, as we have shown, their difference actually decreases with increasing $r_0$. 
We thus conclude that the non-hydrogenic shape plays a role just as important as the average electron-hole separation when computing the interaction strength, thus highlighting the necessity of using the exact wavefunctions in the calculation of the exciton-exciton interaction strength.

\begin{figure}
    \centering
    \includegraphics[width=\columnwidth]{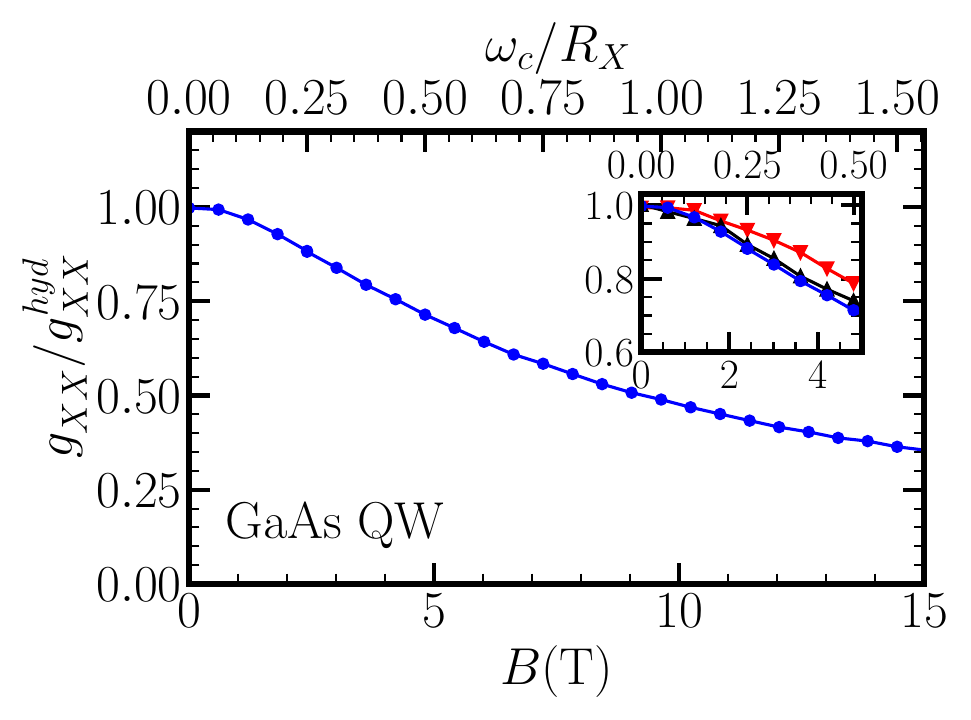}
    \caption{%
    Magnetic field dependence of the exciton-exciton interaction strength \eqref{eq:gXX} for a GaAs QW. 
    In the inset, the small field behavior of our results (blue circles) is compared with the interaction evaluated using the hydrogenic wavefunction $\varphi_{1s}^{hyd} (r)$~\eqref{eq:hydrogen_energy} (red down-triangles), as well as a small-field perturbative expansion to first order of the wavefunction~\eqref{eq:wf_perturbative_smallB} (black upper-triangles). The scale of the upper $x$-axis is the rescaled 
    cyclotron frequency.
    The errorbars of the numerical results (due to statistical error of the Monte Carlo method --- see text) are not visible on the scale of the figure. 
Parameters for the GaAs QW are listed in Table~\ref{tab:parameters}.}
\label{fig:gxx-Coulomb-B}
\end{figure}
%
\subsection{Finite magnetic field}
\label{sec:gXX-finiteMF}
For finite values of the magnetic field, we evaluate numerically the exciton-exciton interaction strength~\eqref{eq:gXX} employing a multidimensional Monte Carlo method~\cite{Press_Numerical_recipes2007}.\footnote{We have checked that, in the zero magnetic field limit, the results obtained with Monte Carlo integration and those obtained by using Eq.~\eqref{eq:gXX-zeroB} coincide within the statistical error of the Monte Carlo integration.} 
Because of the large values of the $1s$ binding energy in TMD monolayers, the diamagnetic shift of the $1s$ state is small and varies in the $\mu$eV range for magnetic field values as high as $60$~T; see Sec.~\ref{sec:exc-magnetic-field}. Because of this, we also expect that $g_{XX}^{}$ has a weak dependence on the magnetic field. For example, for WS$_2$ parameters, at the largest magnetic field considered in this work, $B=60$~T, we obtain only a $~6.5\%$ reduction of the interaction strength compared to the $B=0$ case~\eqref{eq:WS2-gXX_Bzero}:
\begin{equation}
    \frac{\left.g_{XX}^{B=60~\text{T}}\right|_{\text{WS}_2}}{\area} = \frac{5.04\pm0.02} {2\mu}\; .
\end{equation}
The error is the statistical error estimate from Monte Carlo integration.
The small decrease of $g_{XX}^{}$ with increasing magnetic field is partially due to the reduction of the exciton size (see Fig.~\ref{fig:binding-rms-osc-exciton}b). As discussed next, the other factor that comes into play resides in the 
exponential term $e^{-i\frac{e}{2c}[\mathbf{B}\times(\r_e'-\r_e)]\cdot(\r_h'-\r_h)}$ in Eq.~\eqref{eq:gpp}, which originates from the Lamb transformation~\eqref{eq:Xoperator} of the exciton state.

Unlike TMDs, GaAs QWs display 
significant changes in $g_{XX}^{}$ for experimentally accessible values of the magnetic field.  
Using the typical parameters listed in Table~\ref{tab:parameters}, we plot 
in Fig.~\ref{fig:gxx-Coulomb-B} the magnetic field dependence of the exciton-exciton interaction strength. Here, we observe that already at $\omega_c=0.5 R_X$ ($B=4.82$~T) the exciton-exciton interaction strength is reduced by $28.5\%$ compared with the value at $B=0$ ($g_{XX}^{hyd}$), while at $\omega_c=1.5 R_X$ ($B=14.45$~T) the reduction is $63.6\%$. Note that, for GaAs QWs, we limit the range of the magnetic field to $15$~T (see footnote~\ref{footnote3}). 
Note also that, for pure Coulomb interactions, the dependence of $g_{XX}^{}/g_{XX}^{hyd}$ on the rescaled cyclotron frequency $\omega_c/R_X$ is universal and independent of the chosen parameters, such as the exciton reduced mass and Rydberg energy.

In order to determine which factor primarily contributes to the  reduced 
exciton-exciton interaction strength with increasing magnetic field, we plot in the inset of Fig.~\ref{fig:gxx-Coulomb-B} the low-field behavior of the interaction strength. Here, we compare our numerically exact evaluation of Eq.~\eqref{eq:gXX} with two approximations: In the first, the interaction  is evaluated using the hydrogenic wavefunction $\varphi_{1s}^{hyd} (r)$~\eqref{eq:hydrogen_energy} so that the sole  effect of the magnetic field is included in the exponential term $e^{-i\frac{e}{2c}[\mathbf{B}\times(\r_e'-\r_e)]\cdot(\r_h'-\r_h)}$. In the second, the interaction is evaluated by using instead the 
low-field perturbative expansion of the wavefunction to first order~\eqref{eq:wf_perturbative_smallB}. Clearly, the three results agree in the  $B\to 0$
limit. Notably, the first approximation already describes a decrease of the interaction strength with the magnetic field, 
indicating that the mere presence of the exponential in the integrand leads to a reduction of the interaction strength. Including perturbatively the effects of the magnetic field in the exciton wavefunction improves the behavior of the interaction strength with respect to the previous result, approaching the exact result when $B \lesssim 2$~T. 

\begin{figure*}
    \centering
    \includegraphics[width=0.8\textwidth]{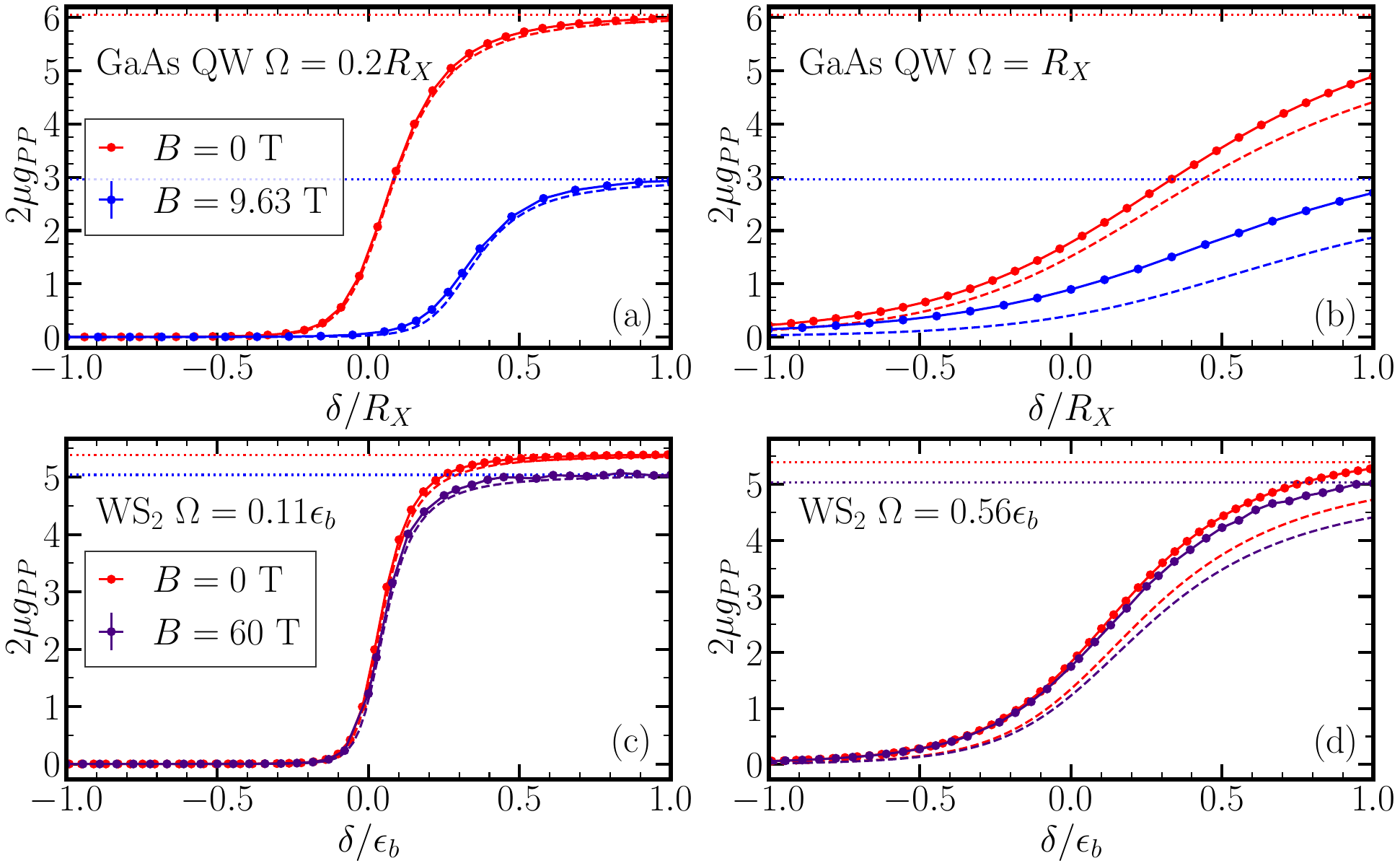}
    \caption{Detuning dependence of the polariton-polariton interaction strength for a GaAs QW (a,b) and an hBN-encapsulated WS$_2$ monolayer (c,d). Parameters are listed in Table~\ref{tab:parameters}. For GaAs QW, $g_{PP}$ is calculated at $B=0$~T and $B=10.72$~T (corresponding to $\omega_c=R_X$),  and for two  
    different values of the  Rabi splitting: (a)  $\Omega=2.7~\text{meV} = 0.2 R_X$; (b) $\Omega= 13.5~\text{meV} = R_X$. For WS$_2$, $g_{PP}$ is calculated at $B=0$~T and $B=60$~T and for two  
    different values of the  Rabi splitting: (c)  $\Omega=20~\text{meV}=0.11 \epsilon_b$; (d) $\Omega= 100~\text{meV}=0.56 \epsilon_b$. Our numerically exact Born approximation (circles) are compared with the perturbative expression~\eqref{eq:gXX-rigidX} (dashed lines). The horizontal dotted lines correspond to the exciton-exciton interaction strength $g_{XX}^{}$ evaluated using Eqs.~\eqref{eq:gXX} and \eqref{eq:gXX-zeroB}. The errorbars due to statistical error of the Monte Carlo method of the numerical results (see text) are not visible on the scale of the figure. }
    \label{fig:gPP}
\end{figure*}
%
\section{Polariton interaction strength}
\label{sec:inter_strength}
We finally examine the interaction properties of magnetopolaritons in 
microcavities containing either
TMD monolayers or
III-V heterostructures.
Interactions between polaritons are inherited from their matter component. Indeed, the hybrid nature of cavity polaritons allows for the possibility of low-mass and strongly interacting quasiparticles. 
Besides, since magnetic fields can be employed to enhance the coupling of exciton states to light~\cite{Pietka_PRB2015,Pietka-Potemski_PRB2017}, we are especially interested in understanding how these interactions are affected by a static transverse magnetic field. 
Recently, it has been proposed that, when there is a large separation of scale between the exciton binding energy and the light-matter coupling, the polariton interaction strength at $B=0$ can be accurately described by a simple analytic expression that only depends on the polariton energy relative to the exciton binding energy~\cite{Bleu2020}, and numerically exact four-body calculations using model electronic potentials have confirmed this result~\cite{Li-Parish-Levinsen_PRB2021}. However, it is not currently known how to extend such approaches to the case of an applied magnetic field, where the internal structure of the exciton is strongly modified. 
Therefore, as for excitons, we employ instead a Born approximation, which has the advantage that it sets an upper bound on the polariton-polariton interaction strength~\cite{Li-Bleu-Levinsen-Parish_PRB2021} for identical excitons (we do not consider the case of polaritons of different spin, for which one needs to go beyond the Born approximation~\cite{Bleu2020}).
Furthermore, the numerically exact four-body calculations in the absence of a magnetic field show that the Born approximation becomes more precise as the Rabi coupling increases~\cite{Li-Parish-Levinsen_PRB2021}.

It was already shown in Ref.~\cite{Levinsen_PRR2019} for pure Coulomb interaction that approaches neglecting the light-induced modifications of the electron-hole wavefunction~\cite{Tassone-Yamamoto_PRB1999,Quattropani_AmerPhysSoc2000,Combescot_EurPhys2007,Ciuti_PRB1998,Glazov_PRB2009} overestimate polariton-polariton interactions in the very strong light-matter coupling regime. We extend the results of Ref.~\cite{Levinsen_PRR2019} to the case of TMD monolayers
and to a finite magnetic field. 
Our approach employs the exact polariton wavefunctions obtained in Sec.~\ref{sec:polaritons}, thus treating 
the coupling to light and the magnetic field on the same footing. 

The derivation of the 
polariton-polariton interaction strength follows the same steps employed in Sec.~\ref{sec:X-inter_strength} for the exciton-exciton interaction strength. In particular, to obtain the Born approximation we consider the uncorrelated two-polariton state:
\begin{equation}
    \ket{P_{\vect{0}}^{2}} = \hat{P}_\0^{\dag}\hat{P}_\0^{\dag} \ket{0} \; ,
\end{equation}
where $\hat{P}_\0^{\dag}$ is the creation operator of a zero momentum polariton defined in Eq.~\eqref{eq:pol-creation-op}. The normalization of this state is identical to Eq.~\eqref{eq:normal_2X_state}, where, however, $\varphi(r)$ is now the wavefunction describing the electron and hole relative motion within the polariton state. $\varphi (r)$ and the photon amplitude $\gamma$ are obtained by solving the coupled equations~\eqref{eq:kappa-space-polaritons} in $\kappa$-space and transforming back to $r$-space. By following the same steps that led to the exciton-exciton interaction strength~\eqref{eq:gXX} in Sec.~\ref{sec:X-inter_strength}, for polaritons we obtain:
\begin{widetext}
\begin{multline}
\label{eq:gpp}
   \frac{g_{PP}^{}}{\area}=-\frac{1}{\mathcal{A}^2}\int d\r_e d \r_h d\r_e'd\r_h'e^{-i\frac{e}{2c}[\mathbf{B}\times(\r_e'-\r_e)]\cdot(\r_h'-\r_h)} 
      \left[V(|\r_e-\r_h|)+V(|\r_e'-\r_h'|)-V(|\r_e-\r_e'|)-V(|\r_h-\r_h'|)\right] \\
      \times \varphi^{*}(|\r_e-\r_h|)\varphi^{*}(|\r_e'-\r_h'|) \varphi(|\r_e'-\r_h|)\varphi(|\r_e-\r_h'|) -\frac{2g\gamma_{ }^{*}}{\area}\int d\r_1 d\r_2 e^{\frac{-ie}{2c}[\mathbf{B}\times \r_1]\cdot\r_2}\varphi(r_1)\varphi(r_2) \varphi^{*}(|\r_1-\r_2|)\; .
\end{multline}
\end{widetext}
The first term of this expression coincides with the exciton-exciton interaction strength~\eqref{eq:gXX}. 
The second term instead derives exclusively from the light-matter coupling and is zero 
in the absence of light-matter coupling, i.e., $g_{XX}^{} = \left.g_{PP}^{}\right|_{\Omega = 0}$.
In the zero magnetic field limit,  the expression~\eqref{eq:gpp} is more conveniently written in momentum space $k$ in which case it recovers the result derived in Ref.~\cite{Levinsen_PRR2019}.

We consider the combined effects of very strong light-matter coupling and a strong magnetic field on the interaction properties between magnetopolaritons for both multiple GaAs quantum wells~\cite{Brodbeck_PRL2017} and multiple WS$_2$ monolayers~\cite{Zhao-Sanvitto_NatComm2023} embedded into a microcavity.
Figure~\ref{fig:gPP} shows our results as a function of detuning. Here we 
compare the results at zero and finite magnetic field, as well as those for two different Rabi splittings corresponding to the strong and very-strong coupling regimes. For GaAs QW, parameters are those characterizing the experiments of Ref.~\cite{Brodbeck_PRL2017}, while for WS$_2$ monolayers, we adopt parameters compatible with those of the experiment of Ref.~\cite{Zhao-Sanvitto_NatComm2023}. In both cases, we compare the numerical results obtained by evaluating Eq.~\eqref{eq:gpp} with the perturbative expression valid in the regime of perturbative light-matter coupling, where the excitonic component within the polariton is assumed to be unmodified by the coupling to light:
\begin{equation}
\label{eq:gXX-rigidX}
    g_{PP}^{(0)}=\beta^4 g_{XX}^{}\; .
\end{equation}
Here, $\beta^2=1- \gamma_{LP}^2$ is the exciton fraction of polaritons valid within the 2-COM~\eqref{eq:2-COM}, where the expression of the Hopfield coefficient $\gamma_{LP}^{}$ is given in Eq.~\eqref{eq:Hope}. As discussed in Sec.~\ref{sec:polaritons}, the perturbative expression~\eqref{eq:gXX-rigidX} is expected to be accurate in the limit $\Omega \ll \epsilon_b$~\cite{Tassone-Yamamoto_PRB1999}.

In both figures, we observe that the polariton-polariton interaction strength interpolates  as a function of detuning between zero and the exciton-exciton interaction strength. However, the behavior deviates from a purely Hopfield factor interpolation as described by Eq.~\eqref{eq:gXX-rigidX} when $\Omega$ is of the order of the $1s$ exciton binding energy. Further, as discussed later, the interpolation is not always monotonic and the polariton-polariton interaction can exceed the exciton-exciton interaction. The approximated expression~\eqref{eq:gXX-rigidX} underestimates the value of the polariton-polariton interaction strength, with deviations from the exact results that are larger at more positive detunings. 
In the very strong coupling regime, when $\Omega$ is of the same order of magnitude as $\epsilon_b$, the discrepancies between exact $g_{PP}^{}$~\eqref{eq:gpp} and perturbative $g_{PP}^{(0)}$~\eqref{eq:gXX-rigidX} increases with increasing magnetic field. This is particularly visible in the case of a GaAs QW in Fig.~\ref{fig:gPP}(b). 
This result is due to the increase of the exciton oscillator strength with increasing magnetic field.  
Consequently,  higher order terms in $\Omega/\epsilon_b$ beyond the leading order approximation $g_{PP}^{(0)}$ are required as the magnetic field grows. Note, however, that it was already shown in Ref.~\cite{Levinsen_PRR2019}, that perturbative corrections due to exciton oscillator saturation~\cite{Tassone-Yamamoto_PRB1999} or photon-assisted exchange processes~\cite{Combescot_EurPhys2007} greatly overestimate the exact result at $B=0$. 

\begin{figure}
    \centering
    \includegraphics[width=\columnwidth]{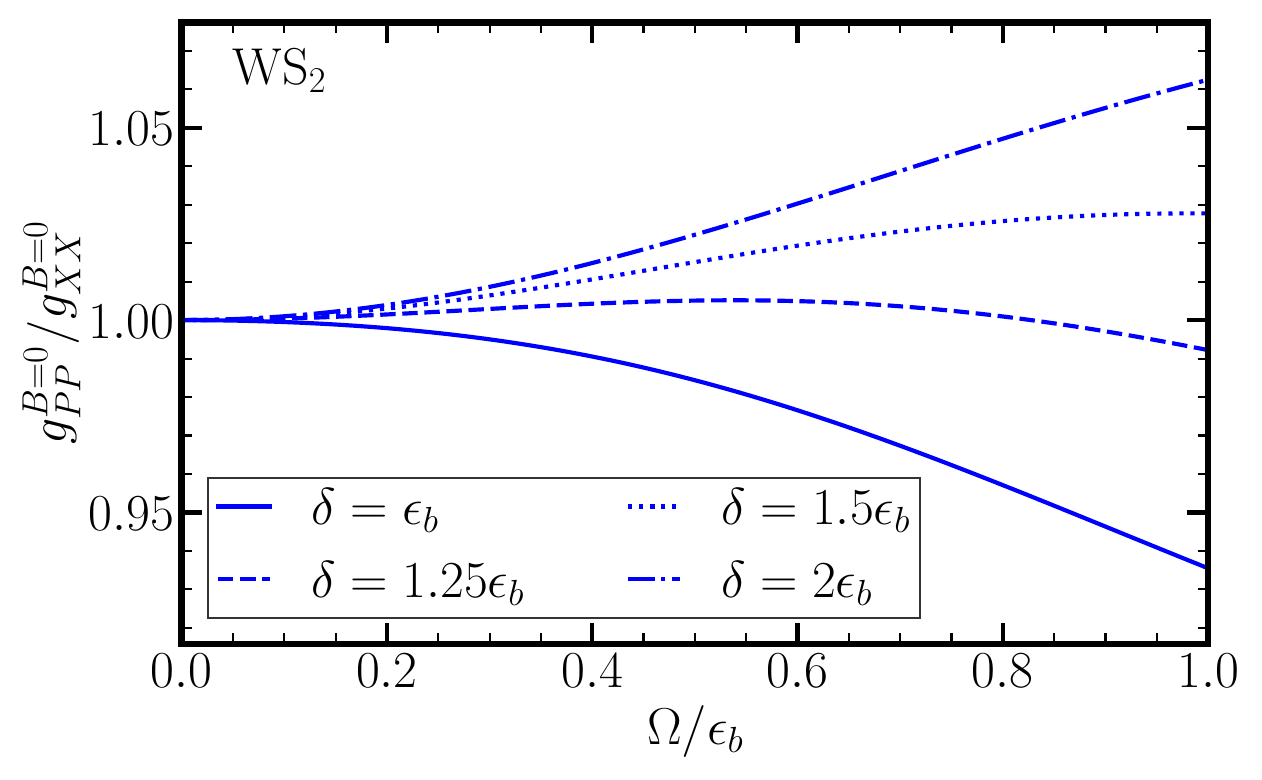}
    \caption{Rabi splitting dependence of the polariton-polariton interaction strength at zero magnetic field for hBN-encapsulated WS$_2$ monolayer for several values of the detuning $\delta$. See system parameters in Table~\ref{tab:parameters}.}
    \label{fig:gpp_omega_dependence}
\end{figure}
The interpolation of the polariton-polariton interaction strength as a function of detuning between zero and the exciton-exciton interaction strength can be slightly non-monotonic and, at positive detunings, $g_{PP}$ can reach values larger than $g_{XX}$. This is illustrated in 
Fig.~\ref{fig:gpp_omega_dependence}, where we plot the dependence on the Rabi splitting of the 
polariton-polariton interaction strength at zero magnetic field and at large positive values of detuning for the specific case of hBN-encapsulated WS$_2$. 
We find that, at sufficiently large detuning, the polariton-polariton interaction strength first grows with $\Omega$ to larger values than the polariton-polariton interaction strength, and then eventually decreases. This implies that $g_{PP}$ can exhibit non-monotonic behavior as a function of detuning and fixed Rabi, approaching at large detunings $g_{XX}$ asymptotically from above rather than from below. This result mirrors those of Refs.~\cite{Bleu2020} and \cite{Li-Parish-Levinsen_PRB2021} which found a similar enhancement of the polariton interactions, with the effect being even larger when one goes beyond the Born approximation.

Finally, we comment on to what extent we can take the polariton interaction strength to be constant in the presence of many polaritons. On general grounds, we expect that as long as the expected polariton blue-shift $g_{PP}n_P$ (with $n_P$ the polariton density) is smaller than the energy scale that controls the interactions, i.e., the light-matter coupling strength~\cite{Bleu2020}, then the many-body effects are small relative to the other relevant scales in the problem. This translates into the condition $g_{PP}n_P\lesssim \Omega/2$. Beyond this regime, one must turn to many-body approaches.

\section{Conclusions}
\label{sec:conclusions}
In this work, we have presented a microscopic approach for TMD monolayers that allowed us to exactly describe excitons and polaritons in the presence of a perpendicular static magnetic field. 
Magneto-optical measurements have already been widely used in TMD monolayers to quantitatively analyze exciton properties and parameters~\cite{Mitioglu_NanoLett2015,Plechinger_NanoLett2016,Stier-Crooker_NanoLett2016,Stier-Crooker_NatComm2016,Stier-Crooker_PRL2018,Zipfel-Chernikov-MF_PRB2018,Have_PRB2019,Liu_PRB2019,Delhomme_APL2019,Goryca-Crooker_NatCom2019}.
Our approach provides numerically exact solutions for the ground and excited states, accommodating arbitrarily large magnetic fields, leading to an extremely good agreement with the diamagnetic shifts measured in Ref.~\cite{Goryca-Crooker_NatCom2019}. 
For polaritons, we have extended our results into the very strong coupling regime, where light-induced modifications of the exciton wavefunction become critical. Here, we have shown that the diamagnetic shift of the lowest-lying polariton  
states at high magnetic fields carries clear signatures of the very strong coupling regime.
Although no current experiments have explored TMD polaritons in this regime, we have investigated experimentally accessible parameters~\cite{Zhao-Sanvitto_NatComm2023}, anticipating that this will inspire future studies.

We have employed the Born approximation to evaluate the interaction properties of both excitons and polaritons in a strong magnetic field, comparing 
TMD structures with  
traditional GaAs quantum wells. In both semiconductor types, we have found that applying a magnetic field leads to a decrease of the interaction strength. 
For TMD excitons, we have shown that employing hydrogenic variational wavefunctions~\cite{Shahnazaryan_PRB_2017} results in overestimating the interaction strength, producing qualitatively inaccurate results and emphasizing the necessity of exact solutions. 
It would be valuable for future research to generalize our formalism to describe the interaction properties of excited states. Rydberg polariton states are gaining significant attention due to their substantial spatial extent,  
which is anticipated to result in significant interactions~\cite{Gu-Menon_NatComm2021, Makhonin-Krizhanovskii_LightSciApp2024}, thus enabling blocking phenomena~\cite{Heckotter_NatComm2021}.

The research data underpinning this publication can be accessed at Ref.~\cite{DATASET}.

\begin{acknowledgments}
We are grateful to 
Scott Crooker and Mateusz Goryca for letting us use the data from Ref.~\cite{Goryca-Crooker_NatCom2019}. We would like to thank Jack Engdahl, Oleg Sushkov, and Dmitry Efimkin for fruitful discussions and for sharing details about their recent unpublished work,  Ref.~\cite{Engdahl_ArXiv2024}. 
DDFP acknowledges financial support from the Ministerio de Educación, Formación Profesional y Deportes through the ``Beca de Colaboración'' fellowship. DDFP and FMM acknowledge financial support from the Spanish Ministry of Science, Innovation and Universities through the ``Maria de Maetzu'' Programme for Units of Excellence in R\&D (CEX2023-001316-M) and from the Comunidad de Madrid and
the Spanish State through the Recovery, Transformation and Resilience Plan
[``Materiales disruptivos bidimensionales 2D'' (MAD2D-
CM)-UAM7]. FMM acknowledges financial support from the Ministry of Science, Innovation and Universities MCIN/AEI/10.13039/501100011033, FEDER UE,  projects No.~PID2020-113415RB-C22 (2DEnLight) and No.~PID2023-150420NB-C31 (Q), and from the Proyecto Sinérgico CAM 2020 Y2020/TCS-6545 (NanoQuCo-CM). MMP is supported through Australian Research Council Future Fellowship FT200100619, and JL through Australian Research Council Discovery Project DP240100569. JL and MMP also acknowledge support from the Australian Research Council Centre of Excellence in Future Low-Energy Electronics and Technologies, `FLEET' (CE170100039).  EL is supported by a Women--in--FLEET research fellowship.
\end{acknowledgments}
\appendix

\section{Two-dimensional hydrogenic solutions and variational approach for $B=0$}
\label{app:Coulomb}
In the limit where the screening length $r_0$ vanishes, the Rytova-Keldysh potential~\eqref{eq:Rytova-Keldysh-potential} recovers the pure Coulomb interaction. In real and momentum space this reads
\begin{align}
\label{eq:coulomb-potential}
    V^{C}(r) &=-\frac{e^2}{\varepsilon r} & V^{C}_{k}&=-\frac{2\pi e^2}{\varepsilon k}\; .
\end{align}
In the absence of a magnetic field, the exciton states recover the 2D hydrogenic solutions~\cite{Yang_PRA1991,Parfitt-Portnoi_JMP2002}. For $s$-wave solutions, the exciton energies are 
\begin{equation}
\label{eq:hydrogen_energy}
    E_{ns}^{hyd} = -\frac{R_X}{(2n-1)^2} \; ,
\end{equation}
and the eigenfunctions:
\begin{equation}
\label{eq:hydrogen_wf}
    \varphi_{ns}^{hyd}(r) = \frac{1}{a_X} \sqrt{\frac{2/\pi}{(2n-1)^3}}\exp(\frac{-r/a_X}{2n-1})\text{L}_{n-1}\left[\frac{2r/a_X}{2n-1}\right]\; ,
\end{equation}
where the Rydberg energy $R_X$ and exciton Bohr radius $a_X$ have been defined in Eq.~\eqref{eq:hydrogen_scales}, and where  $\text{L}_{n-1} (x)$ are the Laguerre polynomials.  
The mean square electron-hole separation can also be calculated analytically:
\begin{equation}
\label{eq:hydrogen_r}
    \langle r^2 \rangle_{ns}^{hyd} = \frac{a_X^2}{2}(2n-1)^2(3+5n(n-1))\; .
\end{equation}

The 2D exciton problem with a Rytova-Keldysh potential~\eqref{eq:Rytova-Keldysh-potential} does not allow for an analytical solution. However, an alternative to solving the \sch equation numerically 
is to consider a variational approach with  trial hydrogenic functions~\eqref{eq:hydrogen_wf}, with the Bohr radius determined variationally for each $ns$ state by minimizing the energy~\cite{Berkelbach_PRB2013,Zipfel-Chernikov-MF_PRB2018,Shahnazaryan_PRB_2017}:
\begin{equation}
    a_X \mapsto \lambda_{X,ns}\; .
\end{equation}
The variational Bohr radii evaluated for the hBN-encapsulated WS$_2$ experiments of Ref.~\cite{Goryca-Crooker_NatCom2019} (see Table~\ref{tab:parameters}) --- and employed in Fig.~\ref{fig:Rydberg-series} to compare the variational and numerically exact exciton energies and radii --- are summarized in Table~\ref{tab:variational_parameters}.
\begin{table}[h]
\begin{tabular}{|l|c|}
\hline
$ns$ & $\lambda_{X,ns}$ (nm) \\ \hline
$1s$ & 1.53          \\ \hline
$2s$ & 0.88          \\ \hline
$3s$ & 0.78          \\ \hline
$4s$ & 0.74          \\ \hline
$5s$ & 0.72          \\ \hline
\end{tabular}
\caption{Variational exciton Bohr radius for the first five $s$-states. Parameters are those summarized in Table~\ref{tab:parameters}, for which $a_X=0.66$~nm. Note that, for growing values of $n$, states become more hydrogenic and $\lambda_{X,ns}\to a_X$.}
\label{tab:variational_parameters}
\end{table}

\section{Subtraction scheme for the Rytova-Keldysh potential}
\label{app:RK-subtraction}
In this appendix we describe the subtraction scheme adopted to numerically solve the \sch equation~\eqref{eq:sch-eq-kappa} that deals with the pole of the Rytova-Keldysh potential in rescaled momentum space, $\tilde{V}_{|\vegr{\kappa} - \vegr{\kappa}'|}$, at $\bm{\kappa}=\bm{\kappa}'$. Note that this method is adapted from the one developed in Ref.~\cite{Laird_PRB2022} for pure Coulomb interaction $|\vegr{\kappa} - \vegr{\kappa}'|^{-1}$. 
The potential $\tilde{V}_{\kappa}$ admits the following analytical expression:
\begin{widetext}
\begin{multline}
\label{eq:RK-kappa-explicit}
    \tilde{V}_{\kappa}=-\frac{1}{9 \bar{r}_0^4 \kappa^{5/2}}\left\{ 128 \sqrt{2} \Gamma \left(\frac{7}{4}\right)^2 \bar{r}_0^2 \kappa \;\; _1F_2\left(1;\frac{5}{4},\frac{5}{4};-\frac{1}{\bar{r}_0^4 \kappa^2}\right)\right. \\
    +8 \sqrt{2} \Gamma\left(\frac{1}{4}\right)^2 \; _1F_2\left(1;\frac{7}{4},\frac{7}{4};-\frac{1}{\bar{r}_0^4 \kappa^2}\right)-9 \pi^2 \bar{r}_0^3  \kappa^{3/2} \left[ Y_0\left(\frac{2}{\bar{r}_0^2 \kappa}\right)+ H_0\left(\left.\frac{2}{\bar{r}_0^2 \kappa}\right)\right]\right\}\; ,
\end{multline}%
\end{widetext}%
where $\bar{r}_0 = r_0/a_X$ and where $_pF_q(a_1,...,a_p;b_1,...,b_q)$ is the generalized hypergeometric function, $H_0(x)$ is the zeroth-order Struve Function and $Y_0(x)$ the zeroth-order Bessel function of the second kind.

In the eigenvalue problem~\eqref{eq:sch-eq-kappa}, the angular integration can be carried out for $s$-wave solutions, which allows us to obtain an equation that only depends on the variable $\kappa$:
\begin{multline}
\label{eq:rew-eigv-proble}
    \bar{E} \int \frac{d\kappa' \kappa'}{2\pi} \tilde{V}_1(\kappa,\kappa') \bar{\varphi}_{\kappa'}\\
    =\kappa^2\bar{\varphi}_{\kappa}+4\bar{\omega}_c^2\bar{\varphi}_{\kappa} + \int \frac{d\kappa' \kappa'}{2\pi} \tilde{V}_2(\kappa,\kappa')\bar{\varphi}_{\kappa'}\;,
\end{multline}
where
\begin{subequations}
\begin{align}
    \tilde{V}_1(\kappa,\kappa') &= \int_0^{2\pi}\frac{d\theta'}{2\pi} \Frac{4\pi}{|\bm{\kappa}-\bm{\kappa}'|} = \Frac{8}{\kappa + \kappa'} \text{K}\left[\frac{4\kappa \kappa'}{(\kappa + \kappa')^2}\right] \\
    \tilde{V}_2(\kappa,\kappa') &= \int_0^{2\pi}\frac{d\theta'}{2\pi} \tilde{V}_{|\bm{\kappa}-\bm{\kappa}'|}\; ,
\end{align}
\end{subequations}
and where $\text{K}(x)$ is the complete elliptic integral of the first kind. While $\tilde{V}_1(\kappa,\kappa')$ can be evaluated analytically~\cite{Laird_PRB2022}, $\tilde{V}_2(\kappa,\kappa')$ has to be evaluated numerically. Note that the potential $\tilde{V}_1(\kappa,\kappa')$ diverges when $\kappa' = \kappa$. However, $\tilde{V}_2(\kappa,\kappa')$ is convergent when $\kappa' = \kappa$ and diverges only when $\kappa' = \kappa = 0$. 
Both issues can be solved by adding and subtracting the following (diagonal) terms to Eq.~\eqref{eq:rew-eigv-proble}:\footnote{Strictly speaking, the subtraction scheme is redundant for the potential $\tilde{V}_2(\kappa,\kappa')$ as we only need to remove the divergence at  $\kappa' = \kappa = 0$. Nevertheless, we find that including it helps with convergence.}
\begin{equation}
\label{eq:convergence}          \int_0^{\infty}\frac{d\kappa'\kappa'}{2\pi} \frac{2\kappa^2}{\kappa^2+\kappa'^2} \left[\bar{E}
        \tilde{V}_{1}(\kappa,\kappa') + \tilde{V}_{2}(\kappa,\kappa')\right] 
\bar{\varphi}_{\kappa}\; .
\end{equation}      
The prefactor $\frac{2\kappa^2}{\kappa^2+\kappa'^2}$ is 1 at $\kappa=\kappa'$, while it decays as $\sim (\kappa')^{-2}$ when $\kappa'\rightarrow\infty$, thus speeding up convergence. At the same time, the terms~\eqref{eq:convergence} that we subtract exactly cancels the divergences of Eq.~\eqref{eq:rew-eigv-proble}, while the terms that we add are convergent. In particular, the following integral can be evaluated analytically:
\begin{equation}
  \int_0^{\infty} \frac{d\kappa'\kappa'}{2\pi}g_1(\kappa,\kappa')= \frac{ \kappa}{\sqrt{2}}\frac{\Gamma\left(\frac{1}{4}\right)^2}{\Gamma\left(\frac{1}{2}\right)}\; ,
\end{equation}
while the other integral can be conveniently rewritten as follows
\begin{equation}
    \int_0^{\infty}\frac{d\kappa'\kappa'}{2\pi}g_2(\kappa,\kappa') = -\frac{2\pi}{\bar{r}_0} \kappa^2 I(\kappa) \; ,
\end{equation}
where
\begin{multline}
\label{eq:I-kappa}
      I(\kappa) = \int_0^{\infty} d\rho J_0(\kappa \rho ) K_0(\kappa \rho)\\ 
      \times \left[H_0\left(\frac{\sqrt{8\rho}}{\bar{r}_0}\right)\notag-Y_0\left(\frac{\sqrt{8\rho}}{\bar{r}_0}\right)\right]\; .
\end{multline}
This integral is evaluated numerically.

\section{Exciton diamagnetic shift for the MoSe$_2$, MoS$_2$, and MoTe$_2$ monolayer data of  Ref.~\cite{Goryca-Crooker_NatCom2019}}
\label{app:other_exp-diamag}
\begin{figure*}
    \centering
    \includegraphics[width=\textwidth]{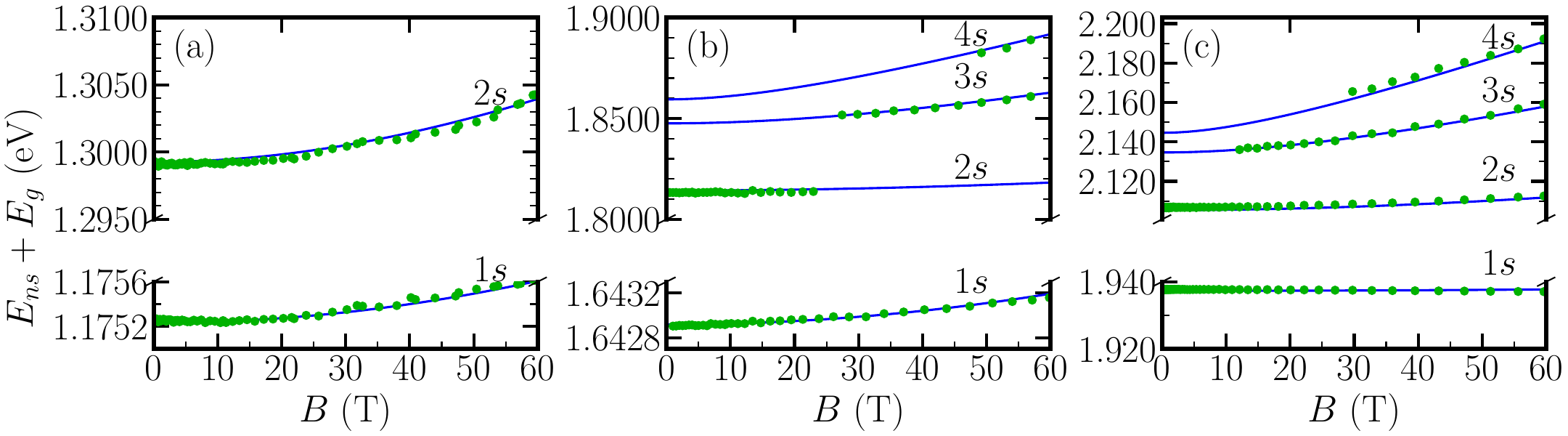}
    \caption{Comparison between theoretical (solid lines) and experimental data of Ref.~\cite{Goryca-Crooker_NatCom2019} (symbols) of the  
    exciton diamagnetic shift of different TMD monolayers. (a) hBN-encapsulated MoTe$_2$ monolayer with parameters $r_0=1.45$~nm, $\mu=0.36 m_0$, $\varepsilon=4.4$, and $E_g=1.35212$~eV. (b) hBN-encapsulated MoSe$_2$ monolayer with parameters $r_0=0.89$~nm, $\mu=0.350 m_0$, $\varepsilon=4.4$, and $E_g=1.87488$~eV. (c) hBN-encapsulated MoS$_2$ monolayer with parameters $r_0=0.76$~nm, $\mu=0.28 m_0$, $\varepsilon=4.5$, and $E_g=2.157$~eV. The same excellent agreement was also shown in Ref.~\cite{Goryca-Crooker_NatCom2019} by using the same parameters, where the \sch equation~\eqref{eq:sch-real-space-exciton} was instead solved in real space.} Note %
    that the energy scale relevant for the $1s$ state is different from that of the excited states to better show variations.
    \label{fig:extra-materials-diamagnetic-shift}
\end{figure*}

In the main text we showed the excellent agreement between our theoretical results and the experimental data of Ref.~\cite{Goryca-Crooker_NatCom2019} on the diamagnetic shift of hBN-encapsulated WS$_2$ monolayer. 
In this appendix we include the comparison for other TMD monolayers in Fig.~\ref{fig:extra-materials-diamagnetic-shift}, namely   MoTe$_2$ (a), MoSe$_2$ (b), and MoS$_2$ (c). The parameters employed to describe these materials have been taken from Ref.~\cite{Goryca-Crooker_NatCom2019} and are indicated in each figure caption, with small variations on the energy bandgap ($<0.1 \%$) for a better comparison. As for the case of WS$_2$, we observe an excellent agreement.


\section{Finite orbital angular momentum exciton states}
\label{app:finiteLz}
We can generalize the results presented in the main text for $s$-wave excitons and obtain the Rydberg series with finite orbital angular momentum.  To this end, we begin with Eq.~\eqref{eq:trans-mat-hamiltonian} and set $\vect{K}=0$, while keeping $l_z$ finite. The exciton \sch equation now reads
\begin{equation}
    E\varphi(\r)=\left[-\frac{1}{2\mu} \nabla_{\r}^2+ \frac{\mu\omega_c^2}{2}r^2+V  (r)\right]\varphi(\r)\; ,
\end{equation}
where $\r = (r,\,\theta)$. We remark that we have neglected the term~\eqref{eq:Lz_1} since $m_e\simeq m_h$ in TMD monolayers, but it is straightforward to include since it simply yields a Zeeman shift of $[(\omega_{c,e}-\omega_{c,h})/2] l_z$ for 
the two states with a given $|l_z| \ne 0$.~\footnote{The effective electron and hole masses for different monolayer TMDs have been calculated from first principles and presented in Table~5 of Ref.~\cite{Rasmussen_PhysChemC2015}.  By using the values for $\mathrm{WS}_2$, the Zeeman shift for $d$-wave excitons ($l_z=\pm2$) at a magnetic field of $B=60~\mathrm{T}$, for example, is around $\pm1.44~\mathrm{meV}$.}

By introducing the rescaled real space variables $\vegr{\rho} = (\rho = r^2/(8a_X^2),\,\phi= 2 \theta)$, we obtain the following dimensionless equation in rescaled real space:
\begin{equation}
\label{eq:rho_finite_Lz}
    \frac{2 \bar{E}}{\rho}\bar{\varphi}(\vegr{\rho})=\left[- \nabla_{\vegr{\rho}}^2+4\bar{\omega}_c^2+\tilde{V} (\rho) \right]\!\bar{\varphi}(\vegr{\rho})\; .
\end{equation}
Fourier transforming this expression into rescaled reciprocal space yields
\begin{equation}
\label{eq:kappa_finite_Lz} \bar{E}\sum_{\bm{\kappa}'}\frac{4\pi\bar{\varphi}_{\vegr{\kappa}'}}{|\bm{\kappa}-\bm{\kappa}'|}= (\kappa^2 +4\bar{\omega}_c^2)\bar{\varphi}_{\vegr{\kappa}}+\sum_{\bm{\kappa}'} \tilde{V} _{|\bm{\kappa}-\bm{\kappa}'|}\bar{\varphi}_{\vegr{\kappa}'}\; .
\end{equation}
Above, $\sum_{\bm{\kappa}}\equiv\int d\bm{\kappa}/(2\pi)^2=\int_0^\infty d\kappa\,\kappa/(2\pi) \int_0^{2\pi} d\phi/(2\pi)$ and $\vegr{\kappa} = (\kappa,\,\phi)$, while $\tilde{V}_\kappa$ is defined in Eq.~\eqref{eq:RK-kappa-explicit}.  
If we now expand the exciton wavefunction $\bar{\varphi}_{\vegr{\kappa}}$ over the orbital angular momentum basis $e^{i\ell \phi}$,
\begin{equation}
\label{eq:partial_wave_expansion}
    \bar{\varphi}_{\vegr{\kappa}} \equiv \bar{\varphi}_{\kappa\phi} = \sum_{\ell\in\mathbb{Z}} e^{i\ell \phi } \tilde{\varphi}_{\kappa\ell}\; ,
\end{equation}
then Eq.~\eqref{eq:kappa_finite_Lz} becomes
\begin{multline}
\label{eq:kappa_ell} 
    \bar{E} \int_0^\infty \frac{d\kappa'\kappa'}{2\pi} V_1 (\kappa,\kappa',\ell) \tilde{\varphi}_{\kappa'\ell}=(\kappa^2+4\bar{\omega}_c^2) \tilde{\varphi}_{\kappa\ell}\\   
    +\int_0^\infty \frac{d\kappa'\kappa'}{2\pi} V_2 (\kappa,\kappa',\ell) \tilde{\varphi}_{\kappa'\ell}\;.
\end{multline}
This result involves the matrix kernels,
\begin{subequations}
\label{eq:kernels_ell} 
\begin{align}
\label{eq:V_1_ell}
    V_1 (\kappa,\kappa',\ell) &= 4\pi\int_0^{2\pi}\frac{d\phi'}{2\pi}\frac{\mathrm{cos}(\ell\phi')}{|\bm\kappa-\bm\kappa'|}\;,\\
\label{eq:V_2_ell}
    V_2 (\kappa,\kappa',\ell) &= \int_0^{2\pi}\frac{d\phi'}{2\pi}\mathrm{cos}(\ell\phi')\tilde{V}_{|\bm\kappa-\bm\kappa'|}\;,
\end{align}
\end{subequations}
where $\phi'$ is the angle of $\bm\kappa'$ measured relative to $\bm\kappa$. 
Note that Eq.~\eqref{eq:kappa_ell} is diagonal in $\ell$ since $[\widehat{H}_m',\,\widehat{L}_z] = 0$, and thus, the equation conserves orbital angular momentum.

Due to the redefinition of the angle $\theta\mapsto\phi=2 \theta$ in the $\r \mapsto \vegr{\rho}$ change of variables, only \textit{even} partial wave states can be accessed by solving the $\bm\kappa$-space equation~\eqref{eq:kappa_ell}: setting $\ell=0,\,\pm 1,\, \pm 2,\,\dots$ gives $l_z=0,\, \pm 2,\,\pm 4,\,\dots$, i.e., the $s,\,d,\,g,\,\dots$ Rydberg exciton series.\footnote{Notice that since the even partial waves are periodic on $2\pi$, the upper limits of the angular integrals in Eqs.~\eqref{eq:kappa_finite_Lz} and~\eqref{eq:kernels_ell} remain as $2\pi$, consistent with the $s$-wave ($l_z=\ell=0$) case.}  Odd partial wave states would instead need to be obtained by solving the Schr\"{o}dinger equation in either real $\r$ space or momentum $\k$ space, both of which are less convenient approaches due to the presence of derivatives.  The subtraction scheme discussed in Ref.~\cite{Laird_PRB2022} and Appendix~\ref{app:RK-subtraction} can be employed to deal with the Coulomb-like pole in the $V_1 (\kappa,\kappa',\ell)$ matrix kernel~\eqref{eq:V_1_ell}.  Numerically integrating over the angle in $V_2 (\kappa,\kappa',\ell)$~\eqref{eq:V_2_ell} can be very time consuming due to the complicated form of the Rytova--Keldysh potential in $\bm\kappa$ space~\eqref{eq:RK-kappa-explicit}.  This process can be sped up significantly first by tabulating the value of $\tilde{V}_\kappa$ versus $\kappa$ on a logarithmic grid and interpolating to populate the three-dimensional matrix kernel, and then by performing Gauss--Legendre quadrature in the angular direction.

\begin{figure}
    \centering
\includegraphics[width=\columnwidth]{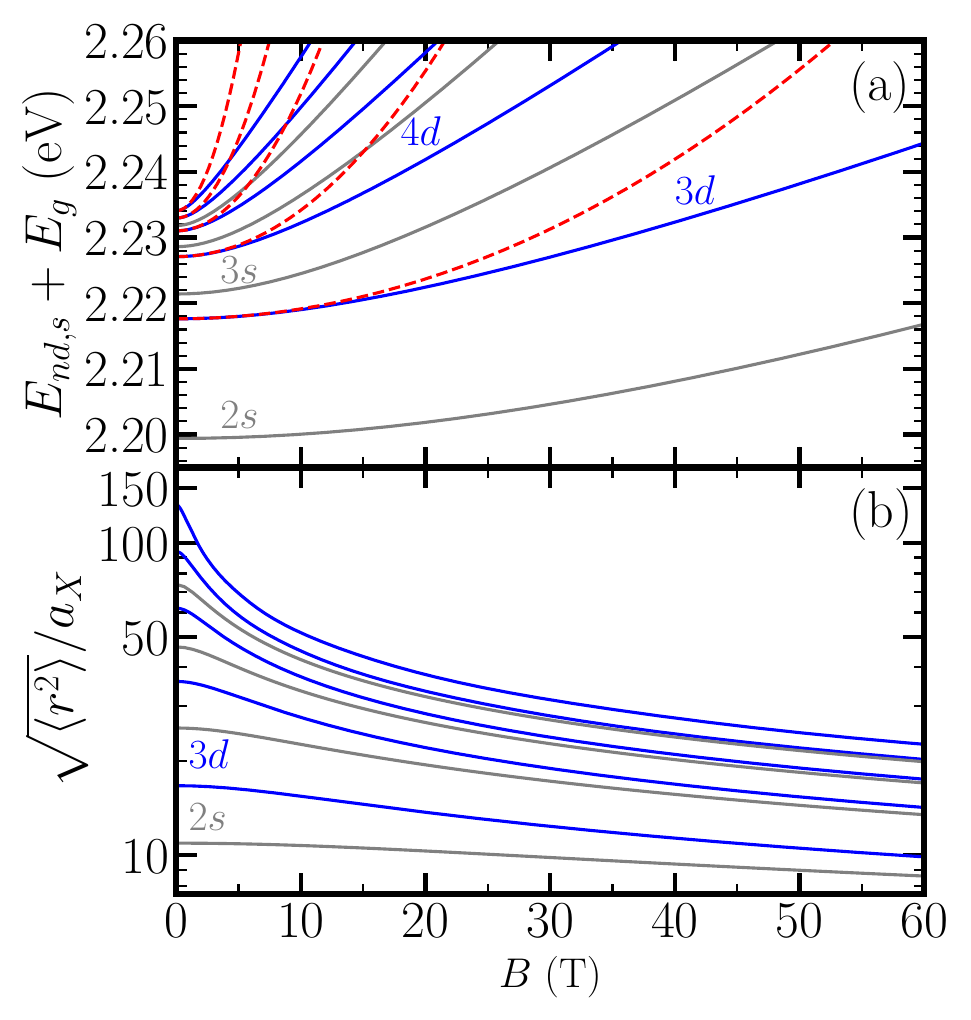}
    \caption{Magnetic field dependence of the first five $nd$ ($n=3,\,\dots,\,7$) Rydberg exciton energies (a) and root-mean-square radii (b)  obtained from the numerically exact solution of Eq.~\eqref{eq:kappa_ell} with $\ell=1$ (solid blue lines). In panel (a) the dashed red lines are the perturbation theory results in the weak magnetic field limit~\eqref{eq:d-wave_weak-field_PT}. In both panels the $ns$ exciton series ($n=2,\,\dots,\,5$) is also shown as a reference (gray lines). All parameters are for hBN-encapsulated WS$_2$ as reported in Ref.~\cite{Goryca-Crooker_NatCom2019} (see Table~\ref{tab:parameters} and $a_X=0.66$~nm), with an energy gap of $E_g=2.23725$~eV.
    }
    \label{fig:d-excitoon-diamagnetic-shift}
\end{figure}
Figure~\ref{fig:d-excitoon-diamagnetic-shift}(a) displays the energies of the Rydberg series of $d$-wave excitons as functions of the magnetic field, overlaid on the $s$-wave exciton energies for comparison.  In the limit where both $B\to0$ and $r_0\to0$, we recover the `accidental degeneracy' of the two-dimensional hydrogen atom~\cite{Parfitt-Portnoi_JMP2002}.  In this case, for each energy level labeled by the principal quantum number $n=1,\,2,\,\dots$, there is a $(2n-1)$-fold degeneracy:
\addtolength{\tabcolsep}{+2pt}
\begin{center}
\begin{tabular}{ r r c }
$n=1$ & 1$s$ & (1-fold) \\ 
2 & 2$s$, 2$p$ & (3-fold) \\  
3 & 3$s$, 3$p$, 3$d$ & (5-fold) \\
4 & 4$s$, 4$p$, 4$d$, 4$f$ & (7-fold)
\end{tabular}
\end{center}
\addtolength{\tabcolsep}{-2pt}
Consistent with Figs.~\ref{fig:Rydberg-series}--\ref{fig:binding-rms-osc-exciton} on magnetoexcitons in the main text, the dimensionless screening length in Fig.~\ref{fig:d-excitoon-diamagnetic-shift}(a) has a finite value of $r_0/a_X\simeq1.18$, which breaks this degeneracy at zero magnetic field, leading to small observable differences between the $3s$ and $3d$ energies, and between the $4s$ and $4d$ energies, etc.  In the opposite limit of strong magnetic fields, the electron-hole interaction potential can effectively be neglected and we instead recover the degeneracies of the Landau levels, i.e., of the two-dimensional harmonic oscillator.  Now, for a given harmonic oscillator index $N=0,\,1,\,2,\,\dots$, there is an $(N+1)$-fold degeneracy:
\addtolength{\tabcolsep}{+2pt}
\begin{center}
\begin{tabular}{ r r c }
$N=0$ & 1$s$ & (1-fold) \\ 
1 & 2$p$ & (2-fold) \\  
2 & 2$s$, 3$d$ & (3-fold) \\
3 & 3$p$, 4$f$ & (4-fold) \\
4 & 3$s$, 4$d$, 5$g$ & (5-fold)
\end{tabular}
\end{center}
\addtolength{\tabcolsep}{-2pt}
Here, the labeling refers to the weak-field states, i.e., how they are adiabatically connected at large $B$.\footnote{We clarify that these high-field degeneracies differ from those in Ref.~\cite{MacDonald-Ritchie_PRB1986}.  There, the authors consider the different problem of a conduction electron interacting with a donor (i.e., an infinitely heavy hole), and therefore the $\mathcal{O}(l_z)$ term~\eqref{eq:Lz_1} is present, which leads to the altered degeneracies shown in their Table III.}  We can see that the parity of $N$ matches whether the angular momentum is even ($s,\,d,\,g,\,\dots$) or odd ($p,\,f,\,\dots$).  Our transformation $\r\mapsto\vegr{\rho}$ 
captures only the even angular momenta, and thus only even $N$, since $N = 2 (n-1) - |l_z|$.  
In the high-field limit, the eigenenergies depend linearly on the exciton cyclotron frequency $\omega_c$.  By plotting and overlaying $E_{ns}/\omega_c$ and $E_{nd}/\omega_c$ versus $\omega_c/R_X$, we have checked that the \{2$s$, 3$d$\} energies approach the same value for increasing $\omega_c/R_X$ ($\gtrsim0.1$), as do the \{3$s$, 4$d$\}, \{4$s$, 5$d$\}, and \{5$s$, 6$d$\} energies.

Also shown in Fig~\ref{fig:d-excitoon-diamagnetic-shift}(a) are the results from first-order non-degenerate perturbation theory in the weak-field regime:
\begin{equation}
\label{eq:d-wave_weak-field_PT}
    E_{nd} \simeq E_{nd}^{B=0} + \frac{\mu\omega_c^2}{2}\langle r^2\rangle_{nd}\; .
\end{equation}
Here, $\langle r^2\rangle_{nd}$ is the mean-square radius of the exciton at zero magnetic field, which is determined by applying the partial wave expansion~\eqref{eq:partial_wave_expansion} both to the following result,
\begin{equation}
   \frac{\langle r^2 \rangle}{a_X^2}=\frac{1}{a_X^2}\int d\r \,r^2 |\varphi(\mathbf{r})|^2=32\sum_{\bm\kappa}|\bar{\varphi}_{\bm\kappa}|^2\, ,
\end{equation}
and to the normalization condition on the wavefunction: $\bar{\varphi}_{\bm\kappa}\rightarrow \bar{\varphi}_{\bm\kappa}/\mathcal{N}$ where $\mathcal{N}^2=8\pi\sum_{\bm{\kappa},\bm{\kappa'}}\bar{\varphi}_{\bm\kappa} \bar{\varphi}_{\bm\kappa'}^{*}/|\bm{\kappa}-\bm{\kappa}'|$.  Similar to the $s$-wave case, for more highly excited states, the perturbative energies are accurate for a narrower range of fields. In other words, as the magnetic field increases, higher excited states approach the strong-field regime more quickly than less excited states.  Figure~\ref{fig:d-excitoon-diamagnetic-shift}(b) displays the root-mean-square electron-hole separation for the $d$-wave Rydberg exciton series.  Again consistent with the $s$-wave case, it can be seen that the exciton size decreases with increasing magnetic field (and increasing binding energy), with the effect more pronounced for higher excited states.  At strong fields, the $s$- and $d$-wave states which tend towards the same radius match the degeneracies in the energy spectrum discussed above.

Exciton states with finite orbital angular momentum are optically dark with regard to conventional one-photon detection~\cite{PhysRevB.62.4927}; however, they may be observed through a consequence of the Stark effect.  By applying a weak static in-plane electric field, one can couple to states that differ by a single quantum of angular momentum~\cite{PhysRevLett.131.036901}.  If the electric field is treated perturbatively, then the leading order element describes the coupling of the $s$ exciton to the $p$ exciton, which causes the latter to shift slightly in energy and become brighter.  The next order element then describes the application of a stronger electric field, whereby the $d$ exciton also couples in and becomes brighter.  Alternatively, excitonic dark states can be probed in TMD monolayers by using two-photon excitation spectroscopy~\cite{Ye_Nature2014}.


%

\end{document}